\documentclass[11pt, onecolumn,  
journal]{IEEEtran}
\usepackage{graphicx,amssymb}
\usepackage{float}
\usepackage{caption}
\usepackage{algorithm}
\usepackage{algorithmic}
\usepackage{amsthm,color}
\usepackage{diagbox}
\usepackage{multirow}
\usepackage{subcaption}
\usepackage{soul}
\usepackage{yhmath}
\usepackage{mathdots}
\usepackage{MnSymbol}
\usepackage{mathrsfs}
\usepackage{lipsum}

\makeatletter

\@addtoreset{algorithm}{section}
\makeatother

\newtheorem{theorem}{Theorem}[section]

\newtheorem{proposition}[theorem]{Proposition}

\newtheorem{corollary}[theorem]{Corollary}

\newtheorem{remark}[theorem]{Remark}

\newtheorem{assumption}[theorem]{Assumption}

\numberwithin{equation}{section}

\def\span{{\rm{span}}}

\setstcolor{red}

\DeclareGraphicsExtensions{.pdf,.jpg,.jpeg,.png}

\begin{document}


\title 
 {Shift-invariant spaces, bandlimited spaces and  reproducing kernel spaces with shift-invariant kernels on undirected finite graphs}

\author{Seok-Young Chung
 and Qiyu Sun
\thanks{Chung and Sun are  with Department of Mathematics, University of Central Florida, Orlando, Florida 32816, USA;
Emails: Seok-Young.Chung@ucf.edu;
 qiyu.sun@ucf.edu.
}}

	\maketitle

\begin{abstract}
 In this paper, we introduce the concept of  graph shift-invariant space (GSIS) on an undirected finite graph, which is the linear space of graph signals being invariant under graph shifts,  and we study its bandlimiting, kernel reproducing and sampling properties.

Graph bandlimited spaces have been widely applied where large datasets on networks need to be handled efficiently.
In this paper, we show that every GSIS is a bandlimited space, and  every bandlimited space is a
  principal GSIS.
  
Functions in a reproducing kernel Hilbert space
with shift-invariant kernel could be learnt with significantly low computational cost.
In this paper, we  demonstrate that every GSIS is a reproducing kernel Hilbert space with a shift-invariant kernel.

Based on the nested  Krylov
structure of GSISs in the spatial domain, we propose a novel sampling and reconstruction algorithm with finite steps, with its  performance
 tested for  well-localized signals on circulant graphs and
flight delay dataset of the  50 busiest airports in the USA.
\end{abstract}


\section{Introduction}

A shift-invariant space  (SIS) $H$ of functions on the  line is a linear space invariant under integer
shifts, i.e.,
$f(\cdot-k)\in H$ for all $f\in H$ and $k\in {\mathbb Z}$.
It has been widely used in approximation theory,   wavelet analysis, sampling theory,
Gabor analysis and many other mathematical and engineering fields
\cite{
 daubechies1992,
   Mallatbook,
   unser2000,  Grochenigbook,
    akram2001, akramsun2001, akramsun2005}.
Bandlimited spaces
and   principal shift-invariant spaces  are two typical examples of SISs.

    Graph signal processing
offers a unique opportunity to  represent, process, analyze, and visualize graph signals and network data
\cite{Shuman2013, Ortega18, Stankovicchapter2019,  Dong2020, Ortegabook, Isufi2023}.
Similar to  the  one-order delay  in classical signal processing,
the concept of  graph shifts has been introduced in graph signal processing.
Graph shifts are usually selected  to have  specific features, and designed to capture the topology of the underlying graph.
Their illustrative examples
include  the adjacency  and  Laplacian matrices of the underlying graph and their variants.
In this paper, we introduce the concept of  graph shift-invariant space (GSIS) on an undirected finite graph,
which is the linear space of graph signals being invariant under graph shifts; see \eqref{sis.def}, and we study  various properties of GSISs on bandlimiting, kernel reproducing and sampling.

Based on  graph Fourier transform on undirected graphs,
bandlimited spaces of graph signals have been well-defined; see  \eqref{bandlimitedspace.def}
and \cite{Pesenson2008, pesenson2009, chen2015, anis2016, 
puy2018, huang2020}.
One may  verify that every graph bandlimited space is invariant under graph shifts.
The first main contribution of this paper is to demonstrate  that
every GSIS is a bandlimited space and  every bandlimited space is a
  principal graph shift-invariant space (PGSIS); see
Theorems \ref{sis.thm} and \ref{finitegenerated.thm}.
Therefore the terminologies on bandlimitedness, shift-invariance and principal-shift-invariance of a linear space of graph signals are essentially the
same in the undirected finite graph setting, with
the first one illustrating  its bandlimiting property
in the Fourier domain, the second one emphasizing its shift-invariance
in the spatial domain, and the third one highlighting the spatial-frequency localization for its generator.

    Reproducing kernel Hilbert spaces  (RKHSs) on the line have been widely accepted in kernel-based learning for  function estimation
\cite{Scholkopf2002, 
  Zhang2009, 
 Shai2014, 
  nikolentzos2021, Chung2023}.
For  efficient learning of functions in a RKHS on an undirected graph,
the kernel are usually selected to be shift-invariant;  see \eqref{siskernel.def}.
RKHSs with  shift-invariant kernels (SIGRKHSs) have the inner product being defined by a generalized dot product in the Fourier domain; see Theorem \ref{rkhs.thm},
and shift-invariant kernels could be learnt with significantly low computational cost.
Common selection of such shift-invariant kernels includes diffusion kernels,  regularization kernels,  random
walk kernels and  spline kernels  \cite{kondor2002, smola2003, zhou2004,  belkin2006,   seto2014, forero2014,   kotzagiannidis2017,  Romero2017,   ward2020,
ncjs22, jian2023}. A SIGRKHS is clearly invariant under graph shifts.
The second main contribution of this paper is to show that the converse is also true;  see Theorem \ref{SisRkhs.thm}.

 SISs and RKHSs on the line
 are suitable for modeling time signals with smoother spectrum, and
 for sampling and numerical implementation of signal reconstruction
\cite{unser2000, akram2001, akramsun2001,  akramsun2005, 
 nashed2010}.
  Flight delays
cause a ripple effect throughout the entire airport network.
  Our numerical simulations indicate that
 the on-time performance data of the 50 busiest airports in the  USA can be 
 modelled as graph signals living
more suitable  in a GSIS with the generators appropriately chosen than in a bandlimited space; see Section \ref{flightdelay.section}.
 We observe that every PGSIS (hence graph bandlimited space and graph shift-invariant space in general)
  has the nested Krylov structure; see \eqref{krylovstructure}.
The third main contribution of this paper is to
establish a sampling theorem for a  PGSIS and propose an iterative
algorithm with finite steps
for graph signal reconstruction,  see Theorem \ref{sampling.thm}, Corollaries  \ref{bandlimitedsampling.cor} and \ref{dynamic.cor}, and Algorithm
\ref{kylovsampling.alg}. Our numerical simulations in Section \ref{numericalsimulation.section} show that the proposed
Algorithm \ref{kylovsampling.alg} has good performance to reconstruct well-localized signals in a  GSIS on circulant graphs and the flight delay  dataset of the 50 busiest airports in the  USA.

The paper is organized as follows. In order to define  GSISs, bandlimited spaces and  SIGRKHSs on undirected graphs,
in Section \ref{preliminary.section}
we recall some preliminaries on graph shifts, polynomial filters, and graph Fourier transform.
In Section \ref{sis.section}, we introduce GSISs, graph bandlimited spaces and GSISs generated by a family of graph signals,
and show that they are essentially the same; see Theorem \ref{sis.thm}.
In Section \ref{sis.section}, we also provide an estimate to Riesz/frame bounds for shifts of the generator of a PGSIS; see
Propositions \ref{Rieszbasis.pr}
and \ref{frame.pr},  and based on the graph uncertainty principle, we show that a PGSIS generated by a localized graph signal has large dimension; see  Proposition \ref{uncertainty.pr}.
In Section \ref{rkhs.section}, we introduce the concept of shift-invariant kernel, discuss the inner product structure of  SIGRKHSs,
 and show that
every GSIS embedded with standard Euclidean inner product is a SIGRKHS; see \eqref{siskernel.def},
and Theorems     \ref{SisRkhs.thm} and  \ref{rkhs.thm}. In Section \ref{sampling.section}, we consider sampling and reconstruction of signals in a PGSIS; see
 Theorem \ref{sampling.thm}, Corollary \ref{bandlimitedsampling.cor} and Algorithm \ref{kylovsampling.alg}. Performance of   Algorithm
  \ref{kylovsampling.alg} for signal reconstruction  is presented in Section \ref{numericalsimulation.section}.
All proofs are collected in Section \ref{proof.section}.

{\bf Notation}: We denote its standard inner product and $p$-norm
 on the
Euclidean space ${\mathbb R}^N$ by
$\langle \cdot, \cdot\rangle$ and $\|\cdot\|_p, 1\le p\le \infty$, respectively.
 We use  $\langle \cdot, \cdot\rangle_H$ and $\|\cdot\|_H$ to represent the inner product and norm on a Hilbert space $H$.
 For a matrix ${\bf A}$, we denote its transpose, maximal singular value, minimal singular value and minimal nonzero singular value by $ {\bf A}^T, \sigma_{\max}({\bf A}), \sigma_{\min}({\bf A})$ and
$\sigma_{\min}^+({\bf A})$ respectively.
As usual, we  denote the set of nonnegative integers by ${\mathbb Z}_+$,
 the cardinality of a set $W$ by  $\#W$, the support of a vector ${\bf x}\in {\mathbb R}^N$ by ${\rm supp}\ {\bf x}$ and
the dimension of a linear space $H$ by $\dim H$.
Also  we set ${\mathbb Z}_+^L=\{[\alpha_1, \ldots, \alpha_L]\ \! | \ \alpha_1, \ldots, \alpha_L\in {\mathbb Z}_+ \}, L\ge 1$
and define
$|{\pmb \alpha}|= \alpha_1+\cdots+\alpha_L$ for ${\pmb \alpha}=[\alpha_1, \ldots, \alpha_L]\in {\mathbb Z}^L_+$.

\section{Preliminaries on graph shifts and graph Fourier transform}
\label{preliminary.section}

Polynomial filters have been widely used in graph signal processing, and graph shifts are building blocks for  polynomial filters
\cite{ Shuman2013, Ortega18,       Isufi2023, ncjs22, aliaksei13, sandryhaila14, jiang19}.
In this section, we first recall some preliminaries on graph shifts and polynomial filters.  
 
 Graph Fourier transform (GFT) decomposes graph signals into different frequency components, and it
    provides a powerful way
 to analyze and process graph signals. Based on multiple commutative graph shifts,  in this section we then
define GFT  
on undirected  graphs \cite{Shuman2013,  Stankovicchapter2019,  Ortegabook, 
Isufi2023, Chung2023,  chung1997, Ricaud2019}.

\subsection{Graph shifts}  Let ${\mathcal G}=(V, E)$ be an undirected graph of order $N\ge 1$.
   We say that  ${\bf S}=[S(i,j)]_{ i,j\in V}$ is a {\em graph shift} if
$S(i,j)=0$ except that either $i= j$  or $(i,j)\in E$.
Illustrative examples of graph  shifts  
are the  adjacency matrix ${\bf A}$, the Laplacian matrix ${\bf L}={\bf D}-{\bf A}$,
 the symmetric	normalized
  Laplacian matrix ${\bf L}^{\rm sym}:={\bf D}^{-1/2} {\bf L}{\bf D}^{-1/2}$,
  and their
	variants, where ${\bf D}$ is the degree matrix.
Similar to  the  one-order delay  in classical multidimensional  signal processing,
the concept of  multiple  commutative graph shifts ${\bf S}_1, \ldots,   {\bf S}_L$ is introduced in \cite{ncjs22}. Here  graph shifts ${\bf S}_1, \ldots,    {\bf S}_L$ are said to be {\em commutative} if
\begin{equation}\label{commutativity.def}
	{\bf S}_l{\bf S}_{l'}={\bf S}_{l'}{\bf S}_l,\  1\le l,l'\le L.
\end{equation}
Under additional real-valued and symmetric assumptions,
they can be diagonalized
simultaneously
by some orthogonal matrix  ${\bf U}$, i.e.,
 \begin{equation}\label{shiftdiagonalization.def} {\bf S}_l= {\bf U} {\pmb \Lambda}_l {\bf U}^T,\  1\le l\le L,\end{equation}
for some diagonal matrices
${\pmb \Lambda}_l={\rm diag}\ \! [\lambda_l(n)]_{1\le n\le N}, 1\le l\le L$ \cite{Chung2023, ncjs22}.
With the help of the  simultaneous diagonalization \eqref{shiftdiagonalization.def},
we define {\em joint spectrum} of commutative graph shifts ${\bf S}_1, \ldots,   {\bf S}_L$ by
	\begin{equation}\label{jointspectrum.def} {\pmb \Lambda}=\big\{{\pmb \lambda}(n)=[\lambda_1(n), ..., \lambda_L(n)]^T\ \!|\ 1\le n\le N\big\}  \subset {\mathbb R}^L.\end{equation}
In this paper,
 we   make the following assumption on the graph shifts  ${\mathbf S}_1, \,\ldots\, ,  {\mathbf S}_L$ and
  their joint spectrum  ${\pmb \Lambda}$.

\begin{assumption}\label{graphshiftassumption}
Graph shifts ${\mathbf S}_1, \,\ldots\, ,  {\mathbf S}_L$ are real-valued, symmetric and commutative, and
elements ${\pmb \lambda}(n), 1\le n\le N$, of  the joint spectrum ${\pmb \Lambda}$ in \eqref{jointspectrum.def} are  distinct.
\end{assumption}

\vskip-0.15in

\subsection{Polynomial filters}
We say that a filter ${\bf H}$ is a {\em polynomial filter} of  graph shifts ${\mathbf S}_1, \ldots, {\bf S}_L$ if
there exists a multivariate polynomial
$h(t_1, \,\ldots\, ,  t_L)=\sum_{[\alpha_1, \ldots, \alpha_L] \in {\mathbb Z}_+^L}
h_{\alpha_1,\ldots, \alpha_L} t_1^{\alpha_1}\cdots t_L^{\alpha_L}$ such that
\begin{equation}\label{MultiShiftPolynomial}
{\bf H}=h({\bf S}_1, \,\ldots\, ,  {\bf S}_L)=\sum_{[\alpha_1, \ldots, \alpha_L]\in {\mathbb Z}_+^L} h_{\alpha_1,\ldots,\alpha_L}{\bf S}_1^{\alpha_1}\cdots {\bf S}_L^{\alpha_L},
\end{equation}
 where the sum is taken on a finite subset of ${\mathbb Z}_+^L$.
 A significant advantage is that the filtering procedure  associated with
 a polynomial filter can be implemented
 at the vertex level in which each vertex is equipped with a one-hop communication subsystem \cite{ncjs22}.
 For a polynomial filter ${\bf H}$, one may verify that it {\em  commutates} with graph shifts ${\bf S}_1, \ldots, {\bf S}_L$, i.e.,
 $ {\bf H} {\bf S}_l={\bf S}_l {\bf H},  1\le l\le L$.
 The converse is shown to  be true in  \cite{aliaksei13} for $L=1$ and \cite[Theorem A.3]{ncjs22} for $L\ge 1$.

\begin{proposition}\label{polynomialfilter.prop} Let ${\mathcal G}$ be a undirected finite graph, and  ${\mathbf S}_1, \ldots, {\mathbf S}_L$ be graph shifts satisfying Assumption \ref{graphshiftassumption}.
Then   ${\bf H}$   commutates with graph shifts ${\bf S}_1, \ldots, {\bf S}_L$
if and only if it is a polynomial filter.
\end{proposition}

By \eqref{shiftdiagonalization.def}  and  Proposition \ref{polynomialfilter.prop}, we see that a filter commutating with commutative graph shifts is diagonalizable.

\begin{corollary}  Let ${\bf U}$ be the orthogonal matrix   in \eqref{shiftdiagonalization.def}
and ${\bf H}$ be a filter  commutating with graph shifts ${\bf S}_1, \ldots, {\bf S}_L$. Then ${\bf U} {\bf H} {\bf U}^T$ is a diagonal matrix.
\end{corollary}

\vskip-0.15in

\subsection{Graph Fourier transform}
For the orthogonal matrix ${\bf U}$ in \eqref{shiftdiagonalization.def}, we write ${\bf U}=[{\bf u}_1, \ldots, {\bf u}_N]$.
 Under Assumption
\ref{graphshiftassumption}, the orthogonal matrix  to  diagonalize  graph shifts ${\bf S}_1, \ldots, {\bf S}_L$ simultaneously
is {\bf unique} up to some sign change, in the sense that, for any orthogonal matrix ${\bf V}=[{\bf v}_1, \ldots, {\bf v}_N]^T$
satisfying ${\bf S}_l= {\bf V} {\pmb \Lambda}_l {\bf V}^T, 1\le l\le L$, there exists
a diagonal matrix ${\bf P}={\rm diag} [ \epsilon_1, \ldots, \epsilon_n]$
for some  $\epsilon_n\in \{-1, 1\}, 1\le n\le N$,  such that
${\bf V}={\bf U} {\bf P}$, or equivalently,
$ {\bf v}_n=\epsilon_n {\bf u}_n$ for all $1\le n\le N$.
In particular, one may verify that  ${\bf P}:={\bf U}^T{\bf V}$  satisfies
${\bf P}{\pmb \Lambda}_l={\pmb \Lambda}_l {\bf P}$ for all $1\le l\le L$,
by the simultaneous diagonalization property for graph shifts ${\bf S}_1, \ldots, {\bf S}_L$.
This together with  Assumption \ref{graphshiftassumption}  implies that ${\bf P}$ is a diagonal matrix.
The  desired conclusion about the diagonal entries of the diagonal matrix ${\bf P}$ then follows from its orthogonality property.

With the uniqueness property of the orthogonal matrix ${\bf U}$  to  diagonalize  graph shifts ${\bf S}_1, \ldots, {\bf S}_L$ simultaneously, we
define the {\em  graph Fourier transform} (GFT)  $\widehat {\bf x}:={\mathcal F} {\bf x}$ of a graph signal ${\bf x}$ by
\begin{equation}
\label{Fourier.def}
\widehat{\bf x}= {\bf U}^T {\bf x}. 
\end{equation}
Analogous to the classical discrete Fourier transform, we may  use interpret
the joint spectrum $\pmb \Lambda$ of graph shifts ${\bf S}_1, \ldots, {\bf S}_L$
    as 
    the set of frequencies to the above  GFT, and their eigenvectors ${\bf u}_1, \ldots, {\bf u}_N$  to form  its graph Fourier basis
\cite{ Shuman2013, Stankovicchapter2019,   Ortegabook,  Isufi2023, Chung2023, chung1997, Ricaud2019, Chen2023, Cheng2023}.

For the  GFT 
 in \eqref{Fourier.def}, we obtain from the orthogonality property of the matrix ${\bf U}$ in \eqref{shiftdiagonalization.def} that
the following Parseval identity
\begin{equation*} 
\|{\mathcal F}{\bf  x}\|_2=\|{\bf x}\|_2\end{equation*}
holds for every graph signal ${\bf x}$.  
By \eqref{shiftdiagonalization.def}, the graph shift operation in the spatial domain is  a multiplier in the Fourier domain.
In particular, for any graph signal ${\bf x}$,  the Fourier transform
of its shifted signal ${\bf S}_l {\bf x}$ is the Hadamard product of
the Fourier transform of the original signal ${\bf x}$ and the vector of eigenvalues of the graph shift ${\bf S}_l$,
\begin{equation} \label{shift.Fourier} \widehat{{\bf S}_l {\bf x}}= {\pmb \Lambda}_l \widehat{\bf x}, \ 1\le l\le L, \end{equation}
 where ${\pmb \Lambda}_l$ are diagonal matrices in \eqref{shiftdiagonalization.def}.

\section{Graph shift-invariant spaces and bandlimited spaces}
\label{sis.section}

Let ${\mathcal G}$ be an undirected graph of order $N\ge 1$,  and ${\bf S}_1, \ldots, {\bf S}_L$ be  commutative graph shifts on ${\mathcal G}$. We say that a linear space $H$ of graph signals
on ${\mathcal G}$ is {\em shift-invariant}
if
\begin{equation}
\label{sis.def} {\mathbf S}_l{\bf x}\in H \ \ {\rm for \ all} \ {\bf x}\in H \ {\rm and} \  1\le l\le L,\end{equation}
and {\em bandlimited}  if
\begin{equation}\label{bandlimit.def}
H= B_\Omega \end{equation}
for some   $\Omega\subset\{1, \ldots, N\}$, where
\begin{equation}\label{bandlimitedspace.def}
B_\Omega=\big\{{\bf x}\ \! | \ {\rm supp} \ \widehat{\bf x}\subset \Omega\big\}.
\end{equation}
The bandlimited spaces $B_\Omega$, also known as Paley-Wiener spaces,  have been widely used in graph signal processing, see
\cite{Pesenson2008, pesenson2009, chen2015, anis2016, 
puy2018, huang2020} 
and references therein.

Let $\Phi$ be a nonempty set of nonzero graph signals on the graph ${\mathcal G}$,  and
set ${\bf S}^{\pmb \alpha}={\bf S}_1^{\alpha_1}\cdots {\bf S}_L^{\alpha_L}$ for ${\pmb \alpha}=[\alpha_1, \ldots, \alpha_L]^T\in {\mathbb Z}_+^L$.
We say that  a graph shift-invariant space (GSIS) $H$
is {\em generated by $\Phi$} if it is the minimal shift-invariant space containing  $\Phi$, i.e.,
\begin{equation}\label{finitegenerated.def0}
H= H(\Phi):= 
{\rm span}\big \{ {\bf S}^{\pmb \alpha} \phi\ \!  |  \  {\pmb \alpha}\in {\mathbb Z}_+^L, 
\ \phi\in \Phi\big\} 
\end{equation}
and that $H$ is {\em principal} if it is generated by a one graph signal.
As $H(\Phi)$ is 
a linear  subspace of ${\mathbb R}^N$, we may assume that 
$\Phi$ 
 has finite cardinality.
By the classical Cayley-Hamilton theorem,
  for every $1\le l\le L$,  ${\bf S}_l^N$ is the linear combination of ${\bf S}_l^m, 0\le m\le N-1$.
 Therefore the  GSIS  generated by  $\Phi$ is given by
\begin{equation} \label{finitegenerated.def}
H(\Phi)={\rm span}  \big\{ {\bf S}_1^{\alpha_1}\cdots {\bf S}_L^{\alpha_L} \phi\ \! |\    0\le  \alpha_1, \ldots, \alpha_L\le N-1, \phi\in \Phi\big\}.
 \end{equation}

 In the classical real-line setting, a bandlimited space is a principal shift-invariant space and a principal shift-invariant space is shift-invariant, while the converse does not hold in general.
In the following theorem, we show that the terminologies on bandlimitedness, shift-invariance, principal-shift-invariance are essentially the same
in the  undirected
finite graph setting.

\begin{theorem} \label{sis.thm} Let ${\mathcal G}$ be an undirected finite graph, and  ${\mathbf S}_1, \ldots, {\mathbf S}_L$ be graph shifts satisfying Assumption \ref{graphshiftassumption}, and $H$ be a linear space of graph signals on the graph ${\mathcal G}$.
Then the following statements are equivalent.
\begin{itemize}

\item[{(i)}] $H$ is a bandlimited space $B_\Omega$ for some $\Omega\subset \{1, \ldots, N\}$.

\item[{(ii)}] $H$ is shift-invariant.

\item[{(iii)}]  $H$ is a finitely-generated shift-invariant space (FGSIS).

\item[{(iv)}] $H$ is a principal shift-invariant space (PGSIS).

\end{itemize}
 \end{theorem}

We divide the proof into the following steps: (iv)$\Longrightarrow$(iii)$\Longrightarrow$(ii)
$\Longrightarrow$(i)
$\Longrightarrow$(iv); see Appendix \ref{sis.thm.pfsection} for the detailed proof.
In the proof of the implication (ii)
$\Longrightarrow$(i), we show that a GSIS $H$ is the bandlimited space $B_\Omega$ with
\begin{equation}\Omega=\cup_{{\bf x}\in H}\ {\rm supp}\ \widehat {\bf x}\  \ {\rm and} \ \ \#\Omega= \dim H,\end{equation}
where the second conclusion on the cardinality of the supporting set $\Omega$ holds as the bandlimited space $B_\Omega$ has dimension $\#\Omega$.

For the case that $H$ is a GSIS   generated by $\Phi$,
we obtain from \eqref{shift.Fourier} that for any $\alpha_1,\ldots, \alpha_L\in {\mathbb Z}_+$ and $\phi\in \Phi$,
 the Fourier transform of  ${\bf S}_1^{\alpha_1} \cdots {\bf S}_L^{\alpha_L} \phi$
has its supported contained in the supporting set of $\widehat \phi$,  i.e.,
$${\rm supp} \ \mathcal F ({\bf S}_1^{\alpha_1} \cdots {\bf S}_L^{\alpha_L} \phi)=
{\rm supp}\ {\pmb \Lambda}_1^{\alpha_1} \cdots {\pmb \Lambda}_L^{\alpha_L}\widehat \phi\subset {\rm supp} \ \widehat \phi.
$$
Therefore the  set $\Omega$ in the bandlimited space $H=B_\Omega$ is completely determined by the generator $\Phi$,
\begin{equation}\label{Omega.Phi}
\Omega= \cup_{\phi\in \Phi}\ {\rm supp}\ \widehat \phi.
\end{equation}

The implication (i)
$\Longrightarrow$(iv) follows from the following theorem, see Appendix \ref{finitegenerated.thm.pfsection} for the  proof.

\begin{theorem} \label{finitegenerated.thm} Let ${\mathcal G}$ be an undirected finite graph, and ${\mathbf S}_1, \ldots, {\mathbf S}_L$ be graph shifts satisfying Assumption \ref{graphshiftassumption}.  Then the bandlimited space $B_\Omega$
is generated  by some graph signal ${\bf \phi}_0\in B_\Omega$
satisfying
\begin{equation}\label{phi0.def}
\widehat \phi_0(n)\ne 0 \ \ {\rm if \ and \ only \ if}\  \ n\in \Omega,\end{equation}
and a graph shift ${\bf T}$, which is
a linear combination of graphs shifts ${\bf S}_1, \ldots, {\bf S}_L$, i.e.,
\begin{equation}\label{finitegenerated.thm.eq1}
B_\Omega 
={\rm span}\big \{ {\bf T}^m { \phi}_0\ \! | \ 0\le m\le \#\Omega-1\big\}.
\end{equation}
 \end{theorem}

We remark that the requirement  \eqref{phi0.def}  for the generator $\phi_0$  is also a necessary condition  for
\eqref{finitegenerated.thm.eq1} to hold.
Let $\chi_\Omega$ be the characteristic function on the set $\Omega$
whose $n$-th component takes value one for every $n\in \Omega$ and value zero for any $n\not\in \Omega$.
Similar to the sinc function for the classical bandlimited space on the real line, our illustrative example of
the generator $\phi_0$ in \eqref{finitegenerated.thm.eq1} for the bandlimited space  $B_\Omega$
 is the  graph signal  ${\mathcal F}^{-1} \chi_\Omega$, the inverse Fourier transform
of the characteristic function $\chi_\Omega$ on the set $\Omega$.

The linear combination requirement for the graph shift ${\bf T}$ in Theorem \ref{finitegenerated.thm}
can be relaxed to the commutativity property between ${\bf T}$ and the graph shifts
${\bf S}_1, \ldots, {\bf S}_L$.
Under the above commutativity assumption,
  there exists
  a diagonal matrix ${\pmb \Lambda}_{\bf T}={\rm diag} [\lambda_{\bf T}(1), \ldots, \lambda_{\bf T}(N)]$
by Assumption
\ref{graphshiftassumption}  such that
\begin{equation}\label{finitegenerated.thm.remeq1}
{\bf T}={\bf U} {\pmb \Lambda}_{\bf T} {\bf U}^T\end{equation}
 where ${\bf U}$ is the orthogonal matrix in \eqref{graphshiftassumption}, see Proposition \ref{polynomialfilter.prop}.
From the proof of Theorem \ref{finitegenerated.thm}, we notice that
a graph shift ${\bf T}$
can be used in  \eqref{finitegenerated.thm.eq1} if and only if
$\lambda_{\bf T}(n), n\in \Omega$, are distinct.

\subsection{Riesz bases for principal shift-invariant spaces} 

For the generator $\phi_0$ and the graph shift ${\bf T}$  of the bandlimited space $B_\Omega$ chosen in \eqref{finitegenerated.thm.eq1}, we define
\begin{equation}\label{vandermonde.def0-}
{\bf V}_{{\bf T}}=\big[(\lambda_{\bf T}(n))^m\big]_{n\in \Omega, 0\le m\le \# \Omega-1}
\end{equation}
and
\begin{equation} \label{vandermonde.def0+}  {\bf V}_{{\bf T}, \phi_0}=
 \big[ \widehat \phi_0(n) (\lambda_{\bf T}(n))^m\big]_{n\in \Omega, 0\le m\le \# \Omega-1},\end{equation}
where 
$\lambda_{\bf T}(n),  n\in \Omega$ are  given in \eqref{finitegenerated.thm.remeq1}, and
 $\widehat \phi_0(n), n\in \Omega$, are the $n$-th component of the Fourier transform $\widehat \phi_0$ of the generator $\phi_0$.
By Theorem \ref{finitegenerated.thm},
$\{{\bf T}^m \phi_0\ \!|\ 0\le m\le \#\Omega-1\}$ is a Riesz basis for the bandlimited space $B_\Omega$.
Applying  Parseval's identity  for the GFT, we have the following Riesz bound estimate.

\begin{proposition}\label{Rieszbasis.pr} Let the graph ${\mathcal G}$, the graph shifts ${\bf S}_1, \ldots, {\bf S}_L$ and ${\bf T}$,
the bandlimited space $B_\Omega$,   and the generator $\phi_0$ be as in Theorem
\ref{finitegenerated.thm}. Then
the minimal
 and maximal singular values 
 of
 the matrix $V_{{\bf T}, \phi_0}$ in \eqref{vandermonde.def0+} are the   low and upper Riesz basis bounds of
 $\{{\bf T}^m \phi_0\ \! |\  0\le m\le \#\Omega-1\}$  of the bandlimited space $B_\Omega$ respectively, i.e.,
\begin{eqnarray}\label{Rieszbasis.cor.eq1}
\sigma_{\min}({\bf V}_{{\bf T}, \phi_0})
\|{\bf c}\|_2 
&\hskip-0.08in  \le &  \hskip-0.08in  \Big\|\sum_{m=0}^{\# \Omega-1} c_m {\bf T}^m \phi_0\Big\|_2 
\le  \sigma_{\max}({\bf V}_{{\bf T}, \phi_0}) \|{\bf c}\|_2 
\end{eqnarray}
hold for all sequences ${\bf c}=[c_m]_{0\le m\le \#\Omega-1}$.
\end{proposition}

We remark  that  ${\bf V}_{{\bf T}, \phi_0}$ in \eqref{vandermonde.def0+} can be considered a weighted version of the Vandermonde matrix ${\bf V}_{\bf T}$
in \eqref{vandermonde.def0-},   and their minimal/maximal singular values are related by
\begin{eqnarray}
& &  \Big(\min_{n\in \Omega}|\widehat \phi_0(n)|\Big)
\sigma_{\min}({\bf V}_{\bf T})\le  \sigma_{\min}({\bf V}_{{\bf T}, \phi_0}) 
\le  \sigma_{\max}({\bf V}_{{\bf T}, \phi_0})\le  \Big(\max_{n\in \Omega}|\widehat \phi_0(n)|\Big)
\sigma_{\max}({\bf V}_{\bf T}).\qquad
\end{eqnarray}
Therefore we may use the minimal/maximal singular values of the Vandermonde matrix ${\bf V}_{\bf T}$
in \eqref{vandermonde.def0-} to estimate lower/upper bounds for the Riesz basis
$\{{\bf T}^m \phi_0\ \!| \ 0\le m\le \#\Omega-1\}$ of the bandlimited space $B_\Omega$.

\subsection{Frames for principal graph shift-invariant spaces}

 As a consequence of Theorems \ref{sis.thm} and \ref{finitegenerated.thm},  every GSIS $H$
  is generated by finite shifts of some generator  $\phi_0$, i.e.,
\begin{equation}\label{finitegenerated.cor.eq1}
H = {\rm span} \big\{ {\bf S}^{\pmb \alpha} \phi_0\ \! | \ {\pmb \alpha}\in \Sigma_{M-1}\big\}
\end{equation}
 hold for all $M\ge \dim H$,
where
$$\Sigma_M=\big\{[\alpha_1, \ldots, \alpha_L]^T\in {\mathbb Z}_+^L\ \!|  \alpha_1+\ldots+\alpha_L\le M \big\}, \  M\ge 1.$$

Unlike one graph shift scenario (i.e., $L=1$),  $\{{\bf S}^{\pmb \alpha} \phi_0 \ |\ {\pmb \alpha}\in \Sigma_{M-1}\}$ with $M\ge \dim H$ is  not
necessarily a basis for the shift-invariant space  $H$ when $L\ge 2$, however by \eqref{shiftdiagonalization.def} and \eqref{finitegenerated.cor.eq1}, we see that
it forms a frame for the GSIS $H$.
Define
\begin{equation} \label{vandermonde.def0}{\bf F}_M:=[{\bf S}^{\pmb \alpha}\phi_0]_{{\pmb \alpha}\in \Sigma_{M-1}}\end{equation}
and  
\begin{equation}\label{vandermonde.def1}
  \widehat {\bf F}_{M, \phi_0}=
 \big[ ({\pmb \lambda}(n))^{\pmb \alpha}  \widehat \phi_0(n) \big]_{1\le n\le N, {\pmb \alpha}\in \Sigma_{M-1}},\end{equation}
where 
 $\widehat \phi_0(n)$ and ${\pmb \lambda}(n), 1\le n\le N$, of the $n$-th component of the Fourier transform $\widehat \phi_0$ of the generator $\phi_0$
 and the joint spectrum $\pmb \Lambda$ of the graph shifts ${\bf S}_1,\ldots, {\bf S}_L$ given  in \eqref{jointspectrum.def},  respectively.
Then one may verify that
 \begin{equation}\label{frame.def0}
{\bf F}_M={\bf U} \widehat {\bf F}_{M, \phi_0},
 \end{equation} where ${\bf U}$ is the orthogonal matrix in \eqref{shiftdiagonalization.def}.
Then applying the Parseval identity  for the GFT, we have the following frame bound estimate. 

\begin{proposition}\label{frame.pr}
  Let the graph ${\mathcal G}$, graph shifts ${\mathbf S}_1, \ldots, {\mathbf S}_L$,
  and the generator $\phi_0$ be as in \eqref{finitegenerated.cor.eq1}. 
   Then
the  minimal nonzero singular value $\sigma_{\min}^+(\widehat{\bf F}_{M, \phi_0})$
 and maximal singular value $\sigma_{\max}(\widehat{\bf F}_{M, \phi_0})$ of
 the matrix $\widehat {\bf F}_{M, \phi_0}$ in \eqref{vandermonde.def1} are the   low and upper frame bounds of the  frame
 $\{{\bf S}^{\pmb \alpha} \phi_0| \ {\pmb \alpha}\in \Sigma_{M-1}\}$ of the shift-invariant space $H$ for all $M\ge \dim H$, i.e.,
\begin{eqnarray}\label{frameboundestimate}
\sigma_{\min}^+(\widehat {\bf F}_{M, \phi_0})
\|{\bf x}\|_2 & \hskip-0.08in \le & \hskip-.08in   \Big (\sum_{{\pmb \alpha}\in \Sigma_{M-1}} |\langle {\bf x}, {\bf S}^\alpha \phi_0\rangle|^2\Big)^{1/2}
\le  \sigma_{\max}(\widehat {\bf F}_{M, \phi_0})
\|f\|_2, \  {\bf x}\in H.\qquad\quad
\end{eqnarray}
\end{proposition}

\subsection{Uncertainty principle  for  principal graph shift-invariant spaces} 

The generator $\phi_0$ selected in \eqref{finitegenerated.cor.eq1}
is more like
the sinc function in the classical Paley-Wiener space.
In this subsection, we  show that a PGSIS generated by a well-localized graph signal does not have small dimension, see
Proposition  \ref{uncertainty.pr}.

For the orthogonal matrix ${\bf U}=[u_n(i)]_{1\le n\le N, i\in V}$  in \eqref{shiftdiagonalization.def},  define
$$\|{\bf U}\|_\infty=\sup_{1\le n\le N, i\in V} |u_n(i)|$$
and
\begin{eqnarray}\label{Uuniformnorm.new}
\|{\bf U}\|_\infty^* &\hskip-0.08in  = &\hskip-0.08in  \sup_{W\subset V, \Omega\subset \{1, \ldots, N\}}
\Big\{ (\# W  \# \Omega )^{-1/2} \ \!\Big|
\sum_{i\in W, n\in \Omega} |u_n(i)|^2\ge 1\Big\}.
\end{eqnarray}
Clearly, we have
\begin{equation}\label{newboundnorm.estimate}
 N^{-1/2} \le \|{\bf U}\|_\infty^*\le \|{\bf U}\|_\infty.
\end{equation}

By the uncertainty principle in \cite{rebrova2023},
\begin{equation}\label{uncertain.eq}
 \#({\rm supp}\ {\bf x})\times \#({\rm supp}\ \widehat {\bf x})\ge \big(\|{\bf U}\|_\infty^*\big)^{-2}
\end{equation}
holds for all nonzero graph signals ${\bf x}$. The reader may refer to \cite{rebrova2023, donoho1989,  
 teke2017}
 and references therein for additional information
on various graph uncertainty principle.

For a nonzero graph signal $\phi_0$, we obtain from
\eqref{Omega.Phi} that the PGSIS generated by $\phi_0$ is a bandlimited space $B_\Omega$ with
$\Omega={\rm supp}\ \widehat \phi_0$. Then we conclude from \eqref{uncertain.eq} and the above observation that
 a PGSIS generated by a localized  graph signal has higher dimension.

\begin{proposition}\label{uncertainty.pr}
 Let the graph ${\mathcal G}$ and the graph shifts ${\mathbf S}_1, \ldots, {\mathbf S}_L$
   be as in Theorem \ref{finitegenerated.thm}. Let $\phi_0$ be a nonzero graph signal with its support denoted by $W$,
   and denote  the PGSIS generated by $\phi_0$ by $H(\phi_0)$. Then
   \begin{equation}\label{uncertainty.preq1}
  \# W \times \dim H(\phi_0) \ge (\|{\bf U}\|_\infty^*)^{-2}
   \end{equation}
where $\|{\bf U}\|_\infty^*$ is given in \eqref{Uuniformnorm.new}.
\end{proposition}

We finish this subsection with a remark on  the optimality of the low bound estimate \eqref{uncertainty.preq1}
 for the PGSIS generated by a localized signal on a circulant graph.

\begin{remark}\label{circulant.remark}  {\rm  Let $1\le q_1<\cdots<q_L<N/2$
such that $q_1, \ldots, q_L, N$ being co-prime,
 and    ${\mathcal C}(N, Q)=(V_N, E_N(Q))$ be  the unweighted circulant graph
generated by  $Q:=\{q_1, \ldots, q_L\}$  that  has the vertex set
$ V_N=\{0, 1, \ldots N-1\}$
and the edge set $E_N(Q)=\{(i, i\pm q \ {\rm mod}\ N), i\in V_N, q\in Q\}$.
Define  ${\bf S}_l=(S_l(i,j))_{i, j\in V_N}, 1\le l\le L$, with $S_l(i, j)=1$ if $j=i$,
$S_l(i,j)=-1/2$ if $i-j=\pm  q_l \ {\rm mod}\  N$ and $S_l(i,j)=0$ otherwise.
One may verify that ${\bf S}_l, 1\le l\le L$, are commutative graph shifts.
Denote the orthogonal matrix  to diagonalize those graph shifts on the circulant graph  ${\mathcal C}(N, Q)$ by ${\bf U}_{N, Q}$.
For the above orthogonal matrix, we have
\begin{equation} \label{circulant.remarkeq1}
 N^{-1/2} \le \|{\bf U}_{N, Q}\|_\infty^*\le \|{\bf U}_{N, Q}\|_\infty\le 2^{1/2} N^{-1/2},
\end{equation}
 see \cite[Section 6.1]{Chen2023}. This indicates that
 the order $N^{-1/2}$ in the lower bound estimate in  \eqref{newboundnorm.estimate} is  optimal for the orthogonal matrix ${\bf U}_{N, Q}$.

Let $\phi_0$ be the delta signal supported at vertex $\lfloor N/2\rfloor+1$, where $\lfloor t\rfloor$ is the integral part of a real number $t$.
One may verify that the  GSIS space $H(\phi_0)$
  contains all  graph signals   ${\bf x}=[x(0),\ldots, x(N-1)]^T$
satisfying the following symmetry property:
$$ x(\lfloor N/2 \rfloor-i) = x(\lfloor N/2 \rfloor+i), \ 1\le i\le N-1-\lfloor N/2 \rfloor,$$
and it
has dimension  $\lfloor N/2 \rfloor+1$.
 Therefore, for the  GSIS $H(\phi_0)$ on the circulant graph ${\mathcal C}(N, Q)$,
the estimate in  \eqref{uncertainty.preq1} becomes  {\bf accurate} when $N$ is odd, and
 the difference between the left and right hand sides in \eqref{uncertainty.preq1} is at most one for even $N$.
} \end{remark}

\section{Shift-invariant graph reproducing kernel Hilbert spaces}
\label{rkhs.section}

Let
  ${\bf S}_1, \ldots, {\bf S}_L$ be  graph shifts on an undirected graph ${\mathcal G}$ of order $N$,
and $H$ be a reproducing kernel Hilbert space (RKHS) of graph signals on ${\mathcal G}$. We say that its reproducing kernel ${\bf K}$
 is
{\em shift-invariant} if it commutates with graph shifts ${\bf S}_1, \ldots, {\bf S}_L$:
\begin{equation}\label{siskernel.def} {\bf S}_l{\bf K}={\bf K} {\bf  S}_l,\  1\le l\le L.\end{equation}
RKHSs with shift-invariant kernel  (SIGRKHSs) have been widely used and appreciated in machine learning for function estimation on networks,
and their popularity can be ascribed to their simplicity to represent functions, flexibility to select kernels,
and efficiency to learn functions with low computational costs.
By Proposition \ref{polynomialfilter.prop}, a shift-invariant reproducing kernel ${\bf K}$ is a polynomial of graph shifts
${\mathbf S}_1, \ldots, {\bf S}_L$.
Common selection of shift-invariant kernels
 include
the diffusion kernels
$\exp(\sigma^2 {\bf L}^{\rm sym}/2)$ with $\sigma>0$,
the $p$-step random walk kernels
$(a{\bf I}-{\bf L}^{\rm sym})^{-p}$ with $a> 2$ and $p\ge 1$, the
Laplacian regularization kernels ${\bf I}+ \sigma^2{\bf L}^{\rm sym}$ with $\sigma>0$, and the spline kernels $ (({\bf L}^{\rm sym})^\dag)^\alpha$, where
$({\bf L}^{\rm sym})^\dag$ is  pseudo-inverse of   the symmetrically normalized Laplacian $ {\bf L}^{\rm sym}$   on the graph ${\mathcal G}$
 \cite{Shuman2013, kondor2002, smola2003,  zhou2004, belkin2006, seto2014,   forero2014,  kotzagiannidis2017,  Romero2017, ward2020,   ncjs22, jian2023}.

 For a  SIGRKHS $H$ with a  shift-invariant kernel ${\bf K}$,
 we can express any element in $H$ as a linear combination of the columns of ${\bf K}$, i.e.,
\begin{equation} \label{SisRkhs.def2} H=\big\{{\bf K} {\bf c},\  {\bf c}\in {\mathbb R}^N\big\}.\end{equation}
The RKHS with a  spline kernel $ (({\bf L}^{\rm sym})^\dag)^p, p\ge 1$,
are  known as graph spline space in \cite{ward2020}.
For the case that  all eigenvalues of the symmetrically normalized Laplacian $ {\bf L}^{\rm sym}$ are distinct (i.e.,  Assumption \ref{graphshiftassumption} holds),
one may verify that the RKHS associated with the diffusion kernels, the $p$-step random walk kernels, and  the
Laplacian regularization kernels are the whole Euclidean space ${\mathbb R}^N$ with different inner products embedded.

By  \eqref{siskernel.def} and \eqref{SisRkhs.def2}, we observe that  a SIGRKHS $H$  is a  GSIS.
In the following theorem, we show that the converse holds too,  see Appendix \ref{SisRkhs.thm.proofsec}
for the  proof.

\begin{theorem}\label{SisRkhs.thm}
Let ${\mathcal G}$ be an undirected finite graph, and ${\mathbf S}_1, \ldots, {\mathbf S}_L$ be  graph shifts satisfying Assumption \ref{graphshiftassumption}.
Then a GSIS $H$ embedded with  the standard Euclidean inner product is a  SIGRKHS.
\end{theorem}

From the proof of Theorem \ref{SisRkhs.thm}, we observe that
  the  inner product
 $\langle {\bf x}, {\bf y}\rangle={\bf x}^T {\bf y}$ of two graph signals ${\bf x}$ and ${\bf y}\in H$  can be represented
 in the Fourier domain as follows:
\begin{equation}\label{standardinnerproduct.def}
\langle {\bf x}, {\bf y}\rangle = {\widehat {\bf x}}^T {\pmb \Lambda}_\Omega \widehat{\bf y},
\end{equation}
where $\Omega\subset \{1, \ldots, N\}$ is the set given in Theorem \ref{sis.thm} and ${\pmb \Lambda}_\Omega= {\rm diag} \big[\chi_{\Omega}(1), \ldots, \chi_\Omega(N))\big]$ is the diagonal matrix with
 $\chi_\Omega(n)=1$ for $n\in \Omega$ and $\chi_\Omega(n)=0$ otherwise.
In the following theorem, we show that
the inner product of any SIGRKHS
can be defined by a generalized dot product  in the Fourier domain, see Appendix \ref{rkhs.thm.pfsection} for the proof.

\begin{theorem}\label{rkhs.thm} Let ${\mathcal G}$ be an undirected finite graph, and  ${\mathbf S}_1, \ldots, {\mathbf S}_L$ be  graph shifts satisfying Assumption \ref{graphshiftassumption}.
Then a SIGRKHS  $H$
has its inner product defined by
\begin{equation}\label{rkhs.thm.eq1}
\langle {\bf x}, {\bf y}\rangle_H= {\widehat {\bf x}}^T {\bf B} {\widehat {\bf y}}, \ \  {\bf x}, {\bf y}\in H,\end{equation}
where ${\bf B}$ is a diagonal matrix with nonnegative diagonal entries.
\end{theorem}

Let ${\bf K}$ be the shift-invariant kernel of a SIGRKHS $H$, $\Omega\subset\{1, \ldots, N\}$ be as in Theorem \ref{sis.thm}
and $\mathbf{U}$ be the orthogonal matrix in \eqref{shiftdiagonalization.def}.
By \eqref{siskernel.def} and Proposition  \ref{polynomialfilter.prop}, we observe that
$\mathbf{K}=\mathbf{U} \mathbf{\Lambda}_{\mathbf{K}} \mathbf{U}^T$
for some diagonal matrix ${\pmb \Lambda}_{\mathbf{K}}$  with
its $n$-th entries taking positive value for $n\in \Omega$ and
zero value otherwise.
  From the proof of Theorem \ref{rkhs.thm}, 
   we observe that the pseudo-inverse ${\pmb \Lambda}_{\bf K}^\dag$ of the diagonal matrix ${\pmb \Lambda}_{\bf K}$ can be used in \eqref{rkhs.thm.eq1} to define
    the inner product of the SIGRKHS $H$. In particular, one may verify that
a matrix ${\bf B}$ with nonnegative diagonal entries can be used  in
\eqref{rkhs.thm.eq1} if and only if ${\bf B}-{\pmb \Lambda}_{\bf K}^\dag$ has
$n$-th diagonal entries taking  zero value for all $n\in \Omega$.

Given  a diagonal matrix  ${\bf B}$ with nonnegative diagonal entries,
one may verify that the bandlimited space
$$B_{\Omega_{\bf B}}=\big\{{\bf x}|\ {\rm supp}\ \widehat {\bf x}\subset \Omega_{\bf B}\big\}$$
embedded with the inner product in \eqref{rkhs.thm.eq1}
is a RKHS with the shift-invariant kernel ${\bf U} {\bf B}^\dag {\bf U}^T$,
where $\Omega_{\bf B}$ is the supporting set of diagonal entries of the diagonal matrix ${\bf B}$
and ${\bf B}^\dag$ is the pseudo-inverse of the matrix ${\bf B}$.

Given a reproducing kernel space $H$ with the inner product defined by \eqref{rkhs.thm.eq1}, we see that the evaluation functional  is uniformly  bounded, i.e.,
\begin{equation} \label{rkhs.thm.eq2}
|x(i)|\le \|{\bf x}\|_2=\|\widehat{\bf x}\|_2\le   \Big(\min_{b(n)\ne 0}  b(n)\Big)^{-1/2} \|{\bf x}\|_H,\  i\in V,
\end{equation}
hold for all graph signals ${\bf x}=[x(i)]_{i\in V}\in H$, where ${\bf B}={\rm diag}(b(1), \ldots, b(N))$ is  the diagonal matrix in \eqref{rkhs.thm.eq1}.

\section{Sampling and reconstruction in a finitely-generated graph shift-invariant space}
\label{sampling.section}

Let
  ${\bf S}_1, \ldots, {\bf S}_L$ be  graph shifts on an undirected graph ${\mathcal G}$ of order $N$,
  $H(\Phi)$ be the GSIS generated by $\Phi=\{\phi_0, \ldots, \phi_{R-1}\}$.
In this section, we consider linear sampling and reconstruction problem in the GSIS $H(\Phi)$,
\begin{equation} \label{sampling.operator} S:\  H(\Phi)\ni {\bf x}\longmapsto {\bf A}{\bf x},\end{equation}
where ${\bf A}=[a(s, i)]_{s\in W, i\in V}$ is the sampling matrix.

An illustrative example of the  sampling  scheme \eqref{sampling.operator} is the ideal sampling on
 $W\subset V$, in which the sampling matrix is given by
 \begin{equation}\label{idealsamplingmatrix} {\bf A}_W=[a_W(i,j)]_{i\in W, j\in V}\end{equation}
 where $a_W(i,j)=0$ except that $a_W(i, i)=1, i\in W$.
The above sampling scheme is also known as {\em subset sampling} \cite{Pesenson2008, pesenson2009, chen2015, anis2016, 
 puy2018, huang2020}.
Another illustrative example of the sampling scheme \eqref{sampling.operator}
is {\em dynamic sampling} (also known as {\em aggregation sampling}),
\begin{equation}\label{dynamicsampling.def}
 S_{{\bf D}, i_0}:\  {\bf x}\longmapsto [{\bf x}(i_0), ({\bf D}{\bf x})(i_0), \ldots, ({\bf D}^{K-1} {\bf x})(i_0)]
\end{equation}
where  $i_0\in V, K\ge 1$,   ${\bf D}$ is the state matrix and  ${\bf x}(i_0)$ is $i_0$-th component of a graph signal ${\bf x}$
\cite{huang2020, akram2013,  marques2016, akram2017, huang2022}.

For the sampling procedure in \eqref{sampling.operator}, we have the following characterization on its injectivity. 

\begin{theorem}\label{sampling.thm} Let ${\mathcal G}$ be an undirected finite graph of order $N$,
 ${\mathbf S}_1, \ldots, {\mathbf S}_L$ be graph shifts satisfying Assumption \ref{graphshiftassumption},   $H(\Phi)$ be the GSIS generated by $\Phi=\{\phi_0, \ldots, \phi_{R-1}\}$, and set the frame matrix
${\bf F}_{N}=[{\mathbf S}^{\pmb \alpha}\phi]_{{\pmb \alpha}\in \Sigma_{N-1}, \phi\in \Phi}$.
Then the sampling procedure \eqref{sampling.operator} with sampling matrix ${\bf A}$ is one-to-one if and only if
${\bf F}_N$ and ${\bf A}{\bf F}_N$
have the same rank.
 \end{theorem}

 Similar to \eqref{frameboundestimate} for PGSISs, we have that the columns of ${\bf F}_N$ form a frame for $H(\Phi)$. Hence
 its rank is the same as the dimension of the GSIS $H(\Phi)$ and also the cardinality of the supporting set $\Omega$ of the corresponding bandlimited space in
 Theorem \ref{sis.thm},
 \begin{equation}\label{rank.eq0}
 {\rm rank}({\bf F}_{N})= \dim H(\Phi)= \# \Omega.
 \end{equation}
This together with
the observation that dimension of  the range space of the sampling operator $S$
is the same as the rank of the matrix ${\bf A}{\bf F}_N$ proves the desired equivalence in Theorem \ref{sampling.thm}.

Let $\Omega\subset \{1, \ldots, N\}$ and $\phi_0\in B_\Omega$ be a nonzero signal with its  GFT $\widehat \phi_0$ has
its  $n$-th component taking nonzero value for all $n\in \Omega$.
Then the GSIS $H(\phi_0)$ generated by  $\phi_0$ is the bandlimited space $B_\Omega$, i.e.,
$ H(\phi_0)= {\rm span} \{{\bf u}_n, n\in \Omega\}$,
where ${\bf U}=[{\bf u}_1, \ldots, {\bf u}_N]$ is the orthogonal matrix in \eqref{shiftdiagonalization.def}.
Combining the above observation with
\eqref{rank.eq0} and Theorem  \ref{sampling.thm}, we obtain the following
result on  the uniqueness for subset sampling on bandlimited spaces.

\begin{corollary} \label{bandlimitedsampling.cor}
Let ${\mathcal G}$ be an undirected graph of order $N$,  ${\mathbf S}_1, \ldots, {\mathbf S}_L$ be graph shifts satisfying Assumption \ref{graphshiftassumption},
and
${\bf U}=[u_n(i)]_{1\le n\le N, i\in V}$ be the orthogonal matrix in \eqref{shiftdiagonalization.def}.
Then for any $\Omega\subset \{1, \ldots, N\}$ and $W\subset V$,  the ideal sampling of the bandlimited space $B_\Omega$ on the sampling set $W$ is one-to-one if and only if
the submatrix $[u_n(i)]_{n\in \Omega, i\in W}$ of the orthogonal matrix ${\bf U}$ has rank $\# \Omega$.
\end{corollary}

The conclusion in Corollary \ref{bandlimitedsampling.cor} has been established in  \cite{Pesenson2008,  chen2015, huang2020} and references therein for
subset sampling on  bandlimited spaces.

For the dynamic sampling \eqref{dynamicsampling.def} with the state matrix ${\bf D}$ being commutative with graph shifts ${\bf S}_1, \ldots, {\bf S}_L$,
one may verify that the corresponding sampling matrix  is given by
\begin{eqnarray*}{\bf A}_{{\bf D}, i_0} & \hskip-0.08in = & \hskip-0.08in  [(\lambda_{\bf D}(n))^k]_{0\le k\le K-1, 1\le n\le N}
 {\rm diag}[u_1(i_0),\ldots, u_N(i_0)] {\bf U},
\end{eqnarray*}
where  
${\rm diag}[\lambda_{\bf D}(1), \ldots, \lambda_{\bf D}(N)]=
{\bf U}^T{\bf D}{\bf U}$, and $[u_1(i_0), \ldots, u_N(i_0)]$ is the $i_0$-th row of the orthogonal matrix.
This together with  Theorem  \ref{sampling.thm} yields the following
result on  the uniqueness of dynamic sampling scheme  \eqref{dynamicsampling.def}
on a GSIS.

\begin{corollary}\label{dynamic.cor}
Let ${\mathcal G}$ be an undirected graph of order $N$,  ${\mathbf S}_1, \ldots, {\mathbf S}_L$ be graph shifts satisfying Assumption \ref{graphshiftassumption}
and   $H(\Phi)$ be the GSIS generated by a finite family  $\Phi$ of  graph signals, and consider the dynamic sampling  scheme $S_{{\bf D}, i_0}$
on the GSIS $H(\Phi)$ at location $i_0\in V$ with the state matrix ${\bf D}$ commutative with graph shifts.
Then the sampling procedure in \eqref{dynamicsampling.def} is one-to-one
 if and only if $K\ge \# \Omega$,  $\lambda_A(n), n\in \Omega$ are distinct,
 and $u_n(i_0)\ne 0$ for all $n\in \Omega$, where $\Omega\subset \{1, \ldots, N\}$ is given in Theorem \ref{sis.thm}.
\end{corollary}


Next consider signal reconstruction associated with the sampling scheme \eqref{sampling.operator}
on  the GSIS $H(\Phi)$, 
under the assumption that the sampling scheme \eqref{sampling.operator} is one-to-one
and  the given  observation data is  the sampling of some signal ${\bf x}_0\in H(\Phi)$ corrupted by some random/deterministic noise ${\pmb \epsilon}$,
\begin{equation}\label{noisyobservation}
{\bf y}={\bf A} {\bf x}_0+{\pmb \epsilon}.
\end{equation}

Let $\Omega\subset \{1, \ldots, N\}$ be the set in Theorem \ref{sis.thm} so that $H(\Phi)=B_\Omega$ and
set ${\bf U}_\Omega=[{\bf u}_n]_{n\in \Omega}$.
Considering the GSIS $H(\Phi)$ as a bandlimited space, we may use
\begin{equation}\label{reconstruction.formula}
{\bf x}^\sharp_0= {\bf U}_\Omega ({\bf U}_\Omega^T {\bf A}^T {\bf A} {\bf U}_\Omega)^{-1} {\bf U}_\Omega^T {\bf A}^T {\bf y},\end{equation}
a solution of the  minimization problem
$${\bf x}_0^\sharp={\rm arg} \min_{{\bf x}\in H(\Phi)} \|{\bf A}{\bf x}-{\bf y}\|_2,$$
as the reconstructed signal in $H(\Phi)$.  
We remark that ${\bf U}_\Omega^T {\bf A}^T {\bf A} {\bf U}_\Omega$ is invertible  by
the assumption that the sampling scheme \eqref{sampling.operator} is one-to-one and its characterization in Theorem \ref{sampling.thm}.
For the numerical implementation of the reconstruction algorithm
\eqref{reconstruction.formula}, we
observe that
${\bf U}_\Omega^T {\bf A}^T {\bf y}$ is essentially the restriction of the Fourier transform of ${\bf A}^T {\bf y}$
onto the set $\Omega$, and hence
we may consider  the reconstruction formula
 \eqref{reconstruction.formula}  for the sampling scheme \eqref{sampling.operator}
 on a GSIS
 as an  implementation   in the Fourier domain.

For $1\le n\le N$, define
\begin{equation}\label{sis.iterativedef}
H_n(\Phi)= {\rm span} \{ {\bf S}^{\pmb \alpha} \phi\ \! | \ {\pmb \alpha}\in \Sigma_n, \phi\in \Phi\}.
\end{equation}
We observe that the GSIS $H(\Phi)$ generated by $\Phi$ could be approximated successively by  nested  Krylov spaces,
\begin{equation}\label{krylovstructure}
{\rm span} \{\Phi\} = H_0(\Phi)  \subset   H_1(\Phi)\subset \cdots  
\subset 
 H_{N-1}(\Phi)=H(\Phi). 
\end{equation}
This inspires us to propose the following iterative algorithm with finite steps to recover the signal ${\bf x}_0^\sharp$ in \eqref{reconstruction.formula},
where
\begin{equation}\label{samplinginnerproduct.def}
\langle {\bf x}_1, {\bf x}_2\rangle_{\bf A}={\bf x}_1^T {\bf A}^T {\bf A} {\bf x}_2
\ \ {\rm
and} \ \  \|{\bf x}_1\|_{\bf A}= \sqrt{\langle {\bf x}_1, {\bf x}_1\rangle_{\bf A}}
\end{equation} for ${\bf  x}_1, {\bf x}_2\in  {\mathbb R}^N$,
 see
Algorithm \ref{kylovsampling.alg}.

\begin{algorithm}
\caption{Reconstruction algorithm for signals in a FGSIS}
{\bf Input}:   $N$ (order of the underlying graph),  ${\bf S}_1, \ldots, {\bf S}_L$ (commutative graph shifts),  $\phi_0, \ldots, \phi_{R-1}$ (generators of the FGSIS),
${\bf A}$ (the sampling matrix),   ${\bf y}={\bf A}{\bf x}_0+{\bf e}$ (the noisy observation of some signal ${\bf x}_0\in H(\phi_0, \ldots, \phi_{R-1})$), and sampling error threshold $\delta$.

{\bf Initial}: Perform Gram-Schmidt process to $\phi_0, \ldots, \phi_{R-1}$ with respect to the inner product $\langle \cdot, \cdot\rangle_{\bf A}$,  and  obtain an orthonormal basis   ${\bf W}_0$.  Set $d_0=\# {\bf W}_0$, ${\bf W}_0=\{{\bf w}_{1}, \ldots, {\bf w}_{d_0}\}$ and ${\bf W}_{-1}=\emptyset$.
Define     ${\bf x}=\sum_{m=1}^{d_0} \langle {\bf y}, {\bf A} {\bf w}_m\rangle {\bf w}_m$ and
    ${\bf e}= {\bf y}- {\bf A} {\bf x}$.

If $\|{\bf e}\|_2\le \delta$, set ${\bf x}_{\rm out}={\bf x}, {\bf e}_{\rm final}={\bf e}$,  $D=0$, and stop, else  do the following iteration.

{\bf Iteration:}

  for $n=1:N-1$
  \begin{itemize}
  \item [{a)}] Set $ {\bf v}_{l, m'}= {\bf S}_l {\bf w}_{m'}- \sum_{m=1}^{d_{n-1}} \langle {\bf S}_l {\bf w}_{m'},
    {\bf w}_m\rangle_{\bf A} {\bf w}_m$
   for all  $1\le l\le L$ and ${\bf w}_{m'}\in {\bf W}_{n-1}\backslash {\bf W}_{n-2}$.

  \item [{b)}] If ${\bf v}_{l, m'}=0$ for all $1\le l\le L$ and ${\bf w}_{m'}\in {\bf W}_{n-1}\backslash {\bf W}_{n-2}$,  
  then set ${\bf x}_{\rm out}={\bf x}, {\bf e}_{\rm final}={\bf e}$, $D=n-1$ and stop.

  \item [{c)}]
  Else perform Gram-Schmidt process to $\{{\bf v}_{l, m'}\ \!|\  1\le l\le L, {\bf w}_{m'}\in {\bf W}_{n-1}\backslash W_{n-2}\}$ with respect to the inner product $\langle \cdot, \cdot\rangle_{\bf A}$,  and  obtain an orthonormal basis
  $\widetilde {\bf W}_n$.

  \item[{d)}] Set $d_n=d_{n-1}+\# \widetilde {\bf W}_n$,  write $\widetilde {\bf W}_n=\{{\bf w}_{d_{n-1}+1}, \ldots, {\bf w}_{d_n}\}$,  and update ${\bf W}_n={\bf W}_{n-1}\cup \widetilde {\bf W}_n$.

  \item [{e)}] Set      ${\bf z}=\sum_{m=d_{n-1}+1}^{d_n} \langle {\bf e}, {\bf A} {\bf w}_m\rangle {\bf w}_m$,
 ${\bf x}={\bf x}+{\bf z}$ and
    ${\bf e}= {\bf e}- {\bf A} {\bf z}$.

  \item [{f)}]  If $\|{\bf e}\|_2\le \delta$, set ${\bf x}_{\rm out}={\bf x}, {\bf e}_{\rm final}={\bf e}$,  $D=n$ and stop, else continue the loop.

\end{itemize}
end

{\bf  Output}: The reconstruction signal ${\bf x}_{\rm out}$, the sampling error ${\bf e}_{\rm final}=\|{\bf y}-{\bf A} {\bf x}_{\rm out}\|$, and
the degree $D$ of graph shifts  to represent the reconstructed signal ${\bf x}_{\rm out}$.

\label{kylovsampling.alg}
\end{algorithm}

  By the one-to-one property of the sampling scheme \eqref{sampling.operator}, we obtain from Theorem \ref{sampling.thm}
  that
  $\langle \cdot, \cdot\rangle_{\bf A}$ in \eqref{samplinginnerproduct.def} defines an inner product on $H(\Phi)$.
For the iteration section of Algorithm \ref{kylovsampling.alg},  one may verify by induction that
in  step a),
$\langle  {\bf v}_{l, m'}, {\bf w}_m\rangle_{\bf A}=0$
      for all $1\le l\le L, {\bf w}_{m'}\in {\bf W}_{n-1}\backslash {\bf  W}_{n-2}$ and ${\bf w}_m\in {\bf W}_{n-1}$;
       in  step d),  $d_n=\dim H_n (\Phi)$,  $\|{\bf w}_{m}\|_{\bf A}= 1$ if $1\le m\le d_n$,
        $\langle {\bf w}_{m}, {\bf w}_{m'}\rangle_{\bf A}= 1$ for all $1\le m\ne m'\le d_n$, and
\begin{equation}\label{orthonormal.Hn} H_n(\Phi)={\rm span} \{{\bf w}_m, 1\le m\le d_n\};\end{equation}
    and in  step e), the reconstruction signal ${\bf x}$ is the solution of the following minimization
    \begin{equation}
\label{algorithm.nstep}    {\bf x}_n:=
    {\rm arg}\min_{{\bf w}\in {\rm span} \widetilde {\bf W}_n} \|{\bf e}-{\bf A} {\bf w}\|_2 ={\rm arg}\min_{{\bf x}\in H_n(\Phi)} \|{\bf y}-{\bf A} {\bf x}\|_2.
    \end{equation}
As the consequence of the above observations, the output ${\bf x}_{\rm out}\in H(\Phi)$ of Algorithm \ref{kylovsampling.alg}
 is  either   a graph signal to match the noisy data ${\bf y}$ within the threshold $\delta$ in the sense that
$$\|{\bf y}-{\bf A}{\bf x}_{\rm out}\|_2\le \delta,$$
or the graph signal   ${\bf x}_0^\sharp$ in  \eqref{reconstruction.formula} that has least approximation error
$$\|{\bf y}-{\bf A}{\bf x}_{\rm out}\|_2=\min_{{\bf x}\in H(\Phi)} \|{\bf y}-{\bf A}{\bf x}\|_2.$$
Comparing with the reconstruction algorithm
 \eqref{reconstruction.formula} for the sampling scheme \eqref{sampling.operator}, we may consider the proposed Algorithm \ref{kylovsampling.alg}
as  a reconstruction scheme in the  spatial domain.  In the next section, we will demonstrate the performance
of the above  algorithm to reconstruct well-localized signals in the spatial domain.

\begin{remark}{\rm
For the orthonormal basis  ${\bf w}_m, 1\le m\le d_D$,  constructed in Algorithm \ref{kylovsampling.alg}, we obtain from
\eqref{orthonormal.Hn} that for any ${\bf w}_m\in {\bf W}_n\backslash {\bf W}_{n-1}, 0\le n\le D$, there exist polynomials $p_{m, r}, 0\le r\le R-1$, of degree $n$ such that
$${\bf w}_m=\sum_{r=0}^{R-1} p_{m, r}({\bf S}_1, \ldots, {\mathbf S}_L) \phi_r.$$
Under the scenario that the sampling matrix ${\bf A}$ satisfies the requirement that  ${\bf A}^T {\bf A}$ is commutative with the graph shifts, we can reformulate the inner
product $\langle {\bf w}_m, {\bf w}_{m'}\rangle_{\bf A}$ between ${\bf w}_m$ and ${\bf w}_{m'}, 1\le m, m'\le d_D$, in the Fourier domain as follows:
\begin{equation}
\langle {\bf w}_m, {\bf w}_{m'}\rangle_{\bf A}  =  \sum_{k=1}^{N} p_{m, r}({\pmb \lambda}(k)) p_{m', r}({\pmb \lambda}(k)) |\widehat \phi_r(k)|^2 \mu_k,
\end{equation}
where ${\pmb \lambda}(k), 1\le k\le N$, are joint spectrum of the graph shifts ${\bf S}_1, \ldots, {\bf S}_L$ given in \eqref{jointspectrum.def} and
${\bf A}^T{\bf A}= {\bf U} {\pmb \Lambda}_{{\bf A}^T{\bf A}} {\bf U}$
for some diagonal matrix ${\pmb \Lambda}_{{\bf A}^T{\bf A}}$ with diagonal entries $\mu_k, 1\le k\le N$.
Under the additional assumption that there is only one graph shifts (i.e.,  $L=1$), we have $d_n=n+1$ for $0\le n\le D$,
and polynomials
$p_{m, 1}$ of degree $m$ are orthogonal polynomials with respect to the discrete measure
$d\mu=\sum_{k=1}^N |\widehat \phi_r(k)|^2 \mu_k \delta_{\lambda(k)}$ supported on the spectrum of the graph shifts,
\begin{equation}
\langle {\bf w}_m, {\bf w}_{m'}\rangle_{\bf A}=\int_{\mathbb R} p_{m, 1}(t) p_{m', 1}(t) d\mu (t).
\end{equation}
Therefore the polynomials
$p_{m, 1}, 0\le m\le D$, satisfies the following three-term recurrence relation of the form
$$p_{m, 1}(t)= (A_mt+ B_m) p_{m-1, 1}(t)+ C_m p_{m-2, 1}, \ 2\le m\le D-1, $$
for some triple $(A_m, B_m, C_m)$ determined by the discrete measure $d\mu$.
}\end{remark}

\section{Numerical Simulations}\label{numericalsimulation.section}

In this section, we evaluate the performance
of Algorithm \ref{kylovsampling.alg} in consideration of reconstruction problem \eqref{sampling.operator} for 
damped cosine wave signals in
a  GSIS on circulant graphs from their noise samples and  flight delay dataset of the 50 busiest airports in the  USA. 

\subsection{GSIS signal reconstruction on circulant graphs}

Let the undirected circulant graph ${\mathcal C}(N, Q)$ 
and the graph shifts
${\bf S}_l, 1\le l\le L$,
be as
in Remark \ref{circulant.remark}.
 Take  the delta signal $\phi_0$ at the vertex $\lfloor N/2 \rfloor$
 and consider
the GSIS $H(\phi_0)$ generated by  $\phi_0$. 
We observe that   the GSIS $H(\phi_0)$ admits the successive Krylov approximation,   
$${\rm span}\{\phi_0\}=H_0(\phi_0) \subseteq \cdots \subseteq H_{N-1}(\phi_0) = H(\phi_0),$$
where Krylov  spaces $H_n(\phi_0),  n\ge 0$, are given by
\begin{equation}\label{circulantKrylov}
    H_n(\phi_0) = \span\big\{ {\bf S}_1^{\alpha_1}\cdots{\bf S}_L^{\alpha_L} \phi_0\ \! \big|\ [\alpha_1,\ldots,\alpha_L]\in \Sigma_{n} \big\}.\end{equation}
On the circulant graph ${\mathcal C}(N, Q)$ with $N=100$ and
  $Q=\{1, 3\}$, one may verify that
  the above Krylov spaces $H_n(\phi_0)$ in \eqref{circulantKrylov}  have dimension $1$ for $n=0$,  $3n$ for $1\le n\le \lfloor N/6\rfloor=16$,
   $50$ for $n=17$ and $51$ for $n\ge 18$.

 \begin{figure}[t] 
\centering
\includegraphics[width=63mm, height=50mm]{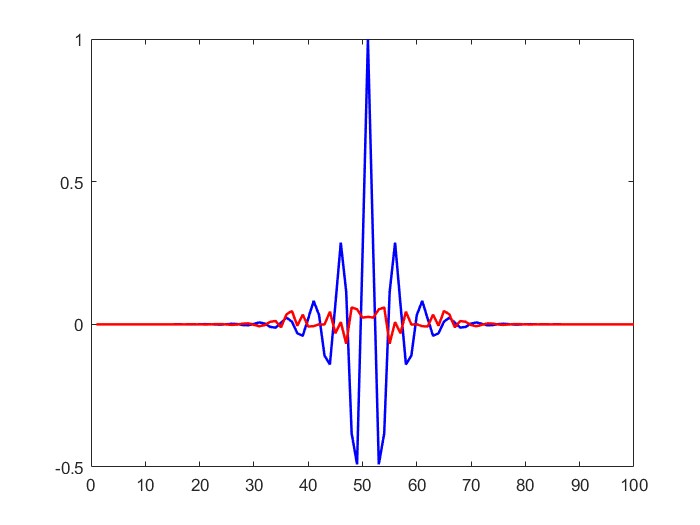}
\includegraphics[width=63mm, height=50mm]{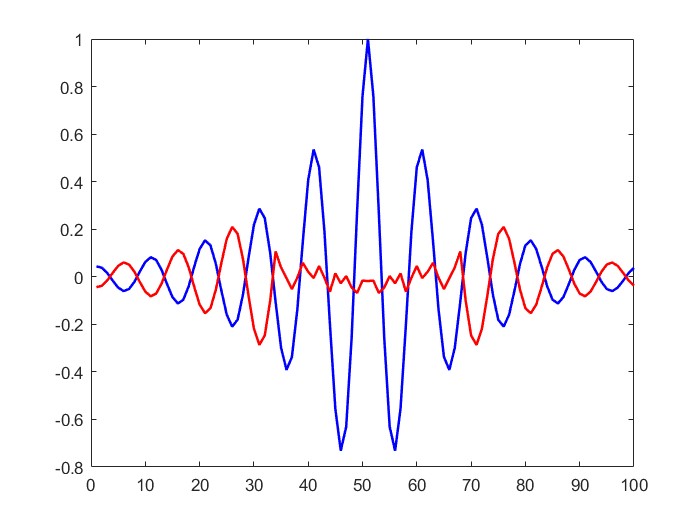}\\
\includegraphics[width=63mm, height=53mm]{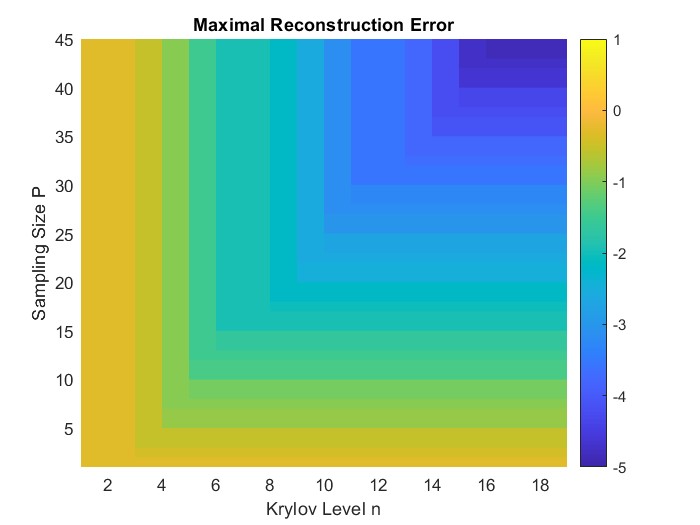}
\includegraphics[width=63mm, height=53mm]{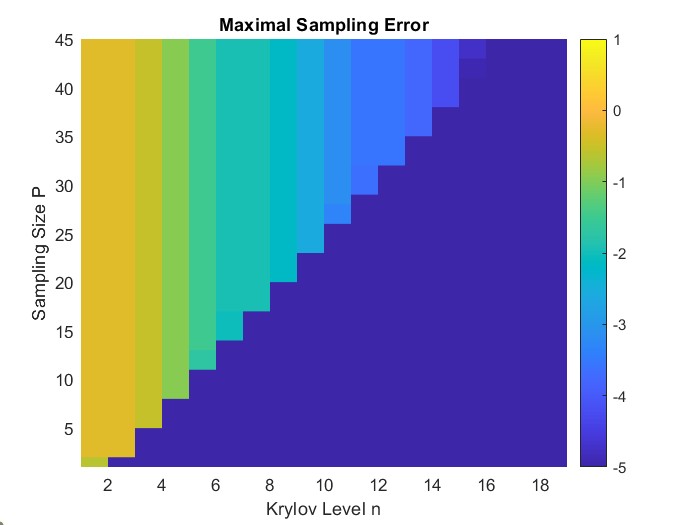}\\
\includegraphics[width=63mm, height=53mm]{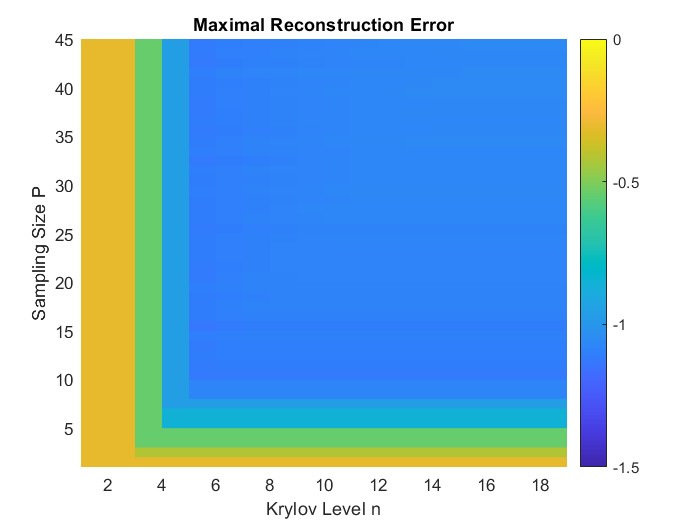}
\includegraphics[width=63mm, height=53mm]{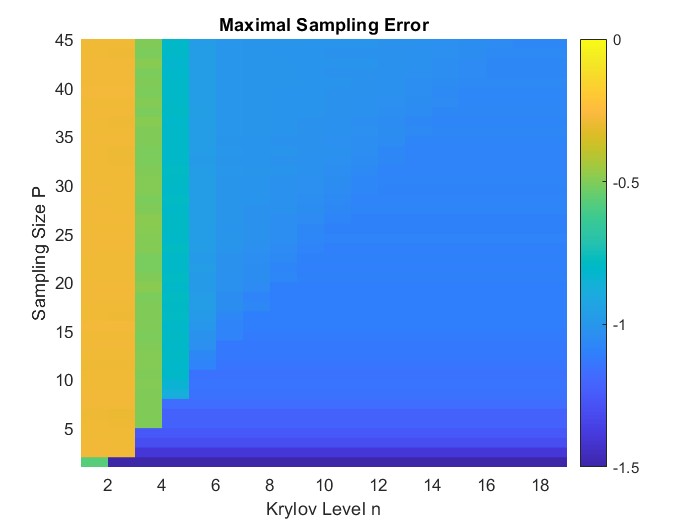}
\caption{Plotted on the top are the damped cosine wave signal ${\bf x}_0$ (in blue)
and the reconstruction error ${\bf x}_{n, W}-{\bf x}_0$ (in red), where 
$A=1, n=6,  N=100,  P=\lfloor N/6\rfloor=16, \sigma=0.1$,  and $(\lambda, \omega)=(1/4, 2\pi/5)$ (top left) and $(1/8, 2\pi/10)$ (top right).
The relative maximal reconstruction error ${\rm RE}(6, 16)$ and the relative maximal sampling error ${\rm SE}(6, 16)$
are $(0.0680, 0.0680)$ (top left)  and $(0.2865, 0.0679)$ (top right) respectively.
Shown in the middle and bottom rows  are the
average of  the relative maximal signal reconstruction error
${\rm RE}(n, P)$ (left) and the relative maximal sampling error ${\rm SE}(n, P)$ (right)
over $M=100$ trials, where $1\le n\le 18, 1\le P\le 45$, ${\bf x}_0$ is the damped cosine wave signal with $A=1, \lambda=1/4$ and $\omega=2\pi/5$,
and the noise level $\sigma= 0$ (the middle row) and $\sigma=0.1$ (the bottom row).
}
\label{SIS.fig}
\end{figure}

 In this subsection, we consider sampling and reconstruction of the damped cosine wave signal  ${\bf x}_0=[x_0(i)]_{0\le i\le N-1}$
 on the circulant graph ${\mathcal C}(N, Q)$,  where
 $A$ is the amplitude, $\lambda$ is the decay constant,  $\omega$ is the angular frequency, and
 \begin{equation}\label{dcws.def} { x}_0(i)=A e^{-\lambda |i-\lfloor N/2 \rfloor|} \cos \big(\omega \big|i-\lfloor N/2 \rfloor\big|\big), \ 0\le i\le N-1.\end{equation}
 The above wave signals belong to the GSIS space $H(\phi_0)$ and are well-localized and symmetric around the vertex $\lfloor N/2 \rfloor$;
  see the top plots of  Figure \ref{SIS.fig} in blue, where $N=100$ and $Q=\{1, 3\}$. 
Moreover, one may verify that maximal relative approximation errors from the Krylov spaces $H_n(\phi_0)$ have exponential decay,
\begin{equation}\label{bestapproximationerror.def}
E_n:=\inf_{{\bf x}_n\in H_n(\phi_0)}\frac{\|{\bf x}_0-{\bf x}_n\|_\infty}{\|{\bf x}_0\|_\infty }\le  e^{-(3n-1)\lambda},\  1\le n\le \lfloor N/6\rfloor.
\end{equation}

  In the simulations, we consider the subset sampling  scheme \eqref{idealsamplingmatrix} with the sampling set
$W$ being symmetric around $\lfloor N/2 \rfloor$,  and the observation data  ${\bf y}$
is corrupted by some uniform random noises,
$${\bf y}(i)={\bf x}_0(i)+\epsilon(i), \ i\in W_P,$$
where
$W_P=\{\lfloor N/2 \rfloor-P, \ldots, \lfloor N/2 \rfloor+P\}$ for some $P\ge 1$,
and $\epsilon(i))_{i\in W_P},$ are i.i.d. random variables with uniform distribution on $[-\sigma, \sigma]$ for some $\sigma>0$.

In the simulations, we use average of  the  relative maximal signal reconstruction error in the logarithmic scale,
$${\rm RE}(n, P)=\log_{10} \Big(\frac{\|{\bf x}_{n, W_P}-{\bf x}_0\|_\infty}{\|{\bf x}_0\|_\infty}+10^{-6}\Big),  $$
and the relative maximal sampling error  in the logarithmic scale,
$${\rm SE}(n, P)=\log_{10} \Big(\frac{\|{\bf A}_{W_P}({\bf x}_{n, W_P}-{\bf x}_0)\|_\infty}{\|{\bf A}_{W_P}{\bf x}_0\|_\infty}+10^{-6}\Big)$$
over $M$ trials
to measure the performance of the Algorithm \ref{kylovsampling.alg},
where  $M\ge 1$, ${\bf A}_{W_P}$ is the sampling matrix, and   ${\bf x}_{n, W_P}\in H_n(\phi_0), n\ge 0$, are  signals reconstructed by Algorithm \ref{kylovsampling.alg}
in its $n$-th iteration; see \eqref{algorithm.nstep}.

Presented in 
 Figure \ref{SIS.fig}
are the performances of Algorithm \ref{kylovsampling.alg} to
reconstruct the damped cosine wave signals
in \eqref{dcws.def} from their noiseless and noisy samples.
For the noiseless scenario shown in the middle row of Figure  \ref{SIS.fig}, we have
the following observations for the relative maximal signal reconstruction error  ${\rm RE}(n, P)$
 and the relative maximal sampling error ${\rm RE}(n, P)$:
 \begin{itemize}
 \item[{1)}]  They
 decrease to zero when Krylov level $n$ and  sampling size $W_P$ increase;
\item[{2)}]  For any given sampling size $1\le P\le \lfloor N/6\rfloor$, they stay unchanged for $n\ge \lfloor (P+1)/3\rfloor+1$; and
 
 \item [{3)}] For any fixed  Krylov level $n\le \lfloor N/3\rfloor$, they take the same value for all $P\ge 3n$.
 \end{itemize}
 Therefore,  for any given sampling size $P$, we may use $n_0=\lfloor P/3\rfloor+1$ as the Krylov level  instead of the default level $N-1$ in Algorithm \ref{kylovsampling.alg}, see the top plots of  Figure \ref{SIS.fig}. Our numerical results also show that
 the maximal relative  signal reconstruction error $\|{\bf x}_{n_0, W_p}-{\bf x}_0\|_\infty/\|{\bf x}_0\|_\infty$  is proportional to the best maximal approximation
 error $E_n$ in \eqref{bestapproximationerror.def}.
The possible reasons we believe are that signals in the Krylov space $H_n(\phi)$ are supported in $[\lfloor N/2\rfloor-3n, \lfloor N/2\rfloor+3n]$
 and they can be fully recovered from their samples on vertices in that interval.

For the noisy scenario shown in the bottom row of Figure   \ref{SIS.fig},
we observe that the  relative maximal signal reconstruction error  ${\rm RE}(n, P)$
 and the relative maximal sampling error ${\rm SE}(n, P)$ have the same monotonic property about the
 Krylov level $n$ and the sampling size $P$, however they do not improve significantly when $P\ge 12$ and $n\ge 5$.
 The possible reason is that the damped cosine wave function ${\bf x}_0$ is well-localized around $\lfloor N/2\rfloor$ with
 $\sup_{|i-\lfloor N/2\rfloor|\ge 13} |x_0(i)|=0.0235$  and its noisy samples at vertices $i$ with
 $|i-\lfloor N/2\rfloor|\ge 13$ are dominated by random noises  at level $\sigma=0.1$.

\subsection{Flight-delay dataset and graph shift-invariant spaces}
\label{flightdelay.section}

\begin{figure}[t] 
\centering
\includegraphics[width=63mm, height=53mm]{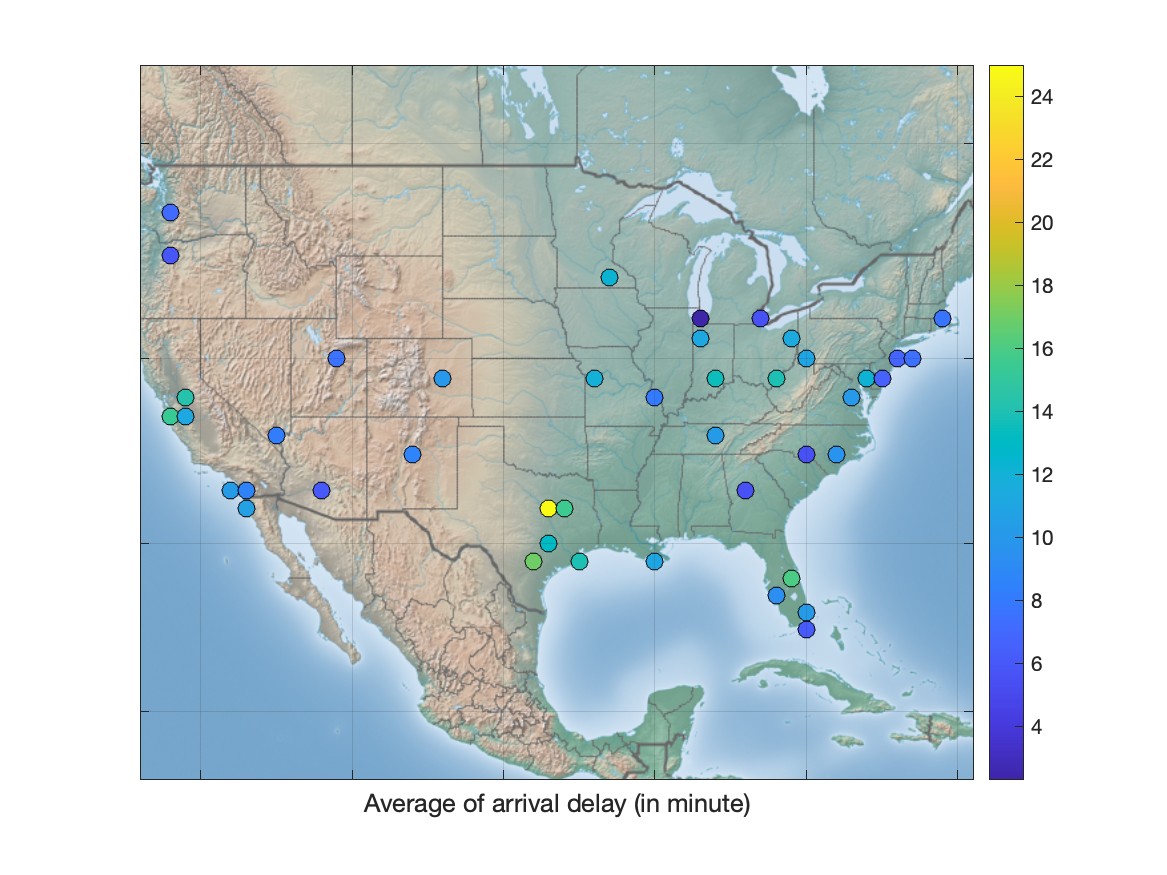}
\includegraphics[width=63mm, height=53mm]{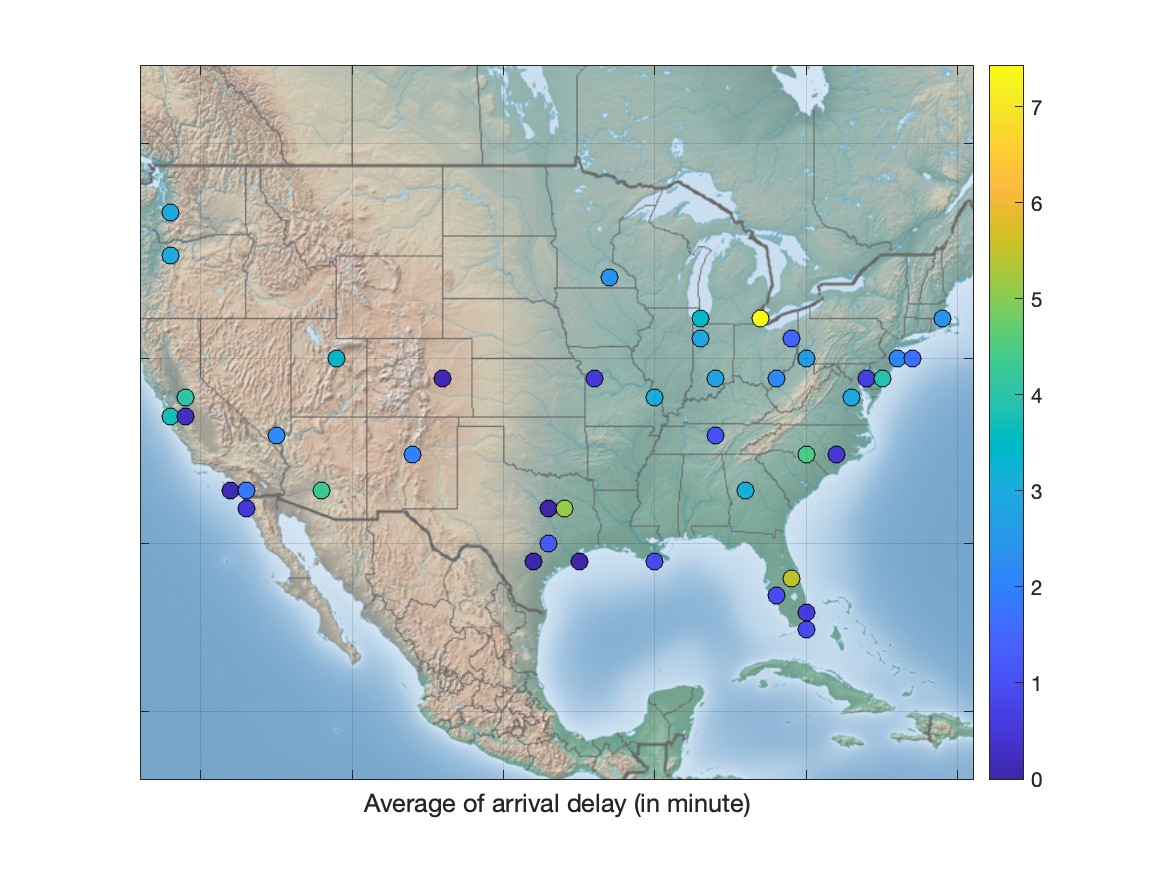}\\
\includegraphics[width=63mm, height=50mm]{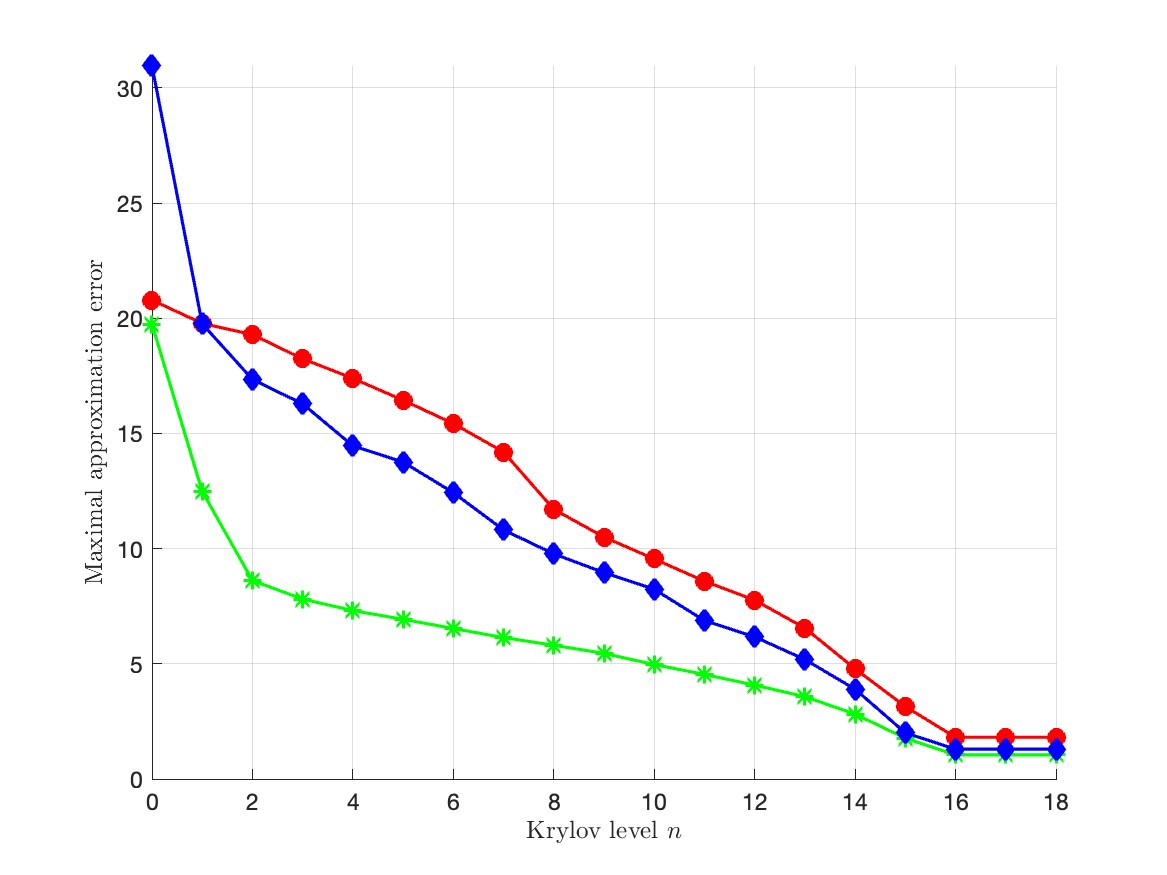}
\includegraphics[width=63mm, height=50mm]{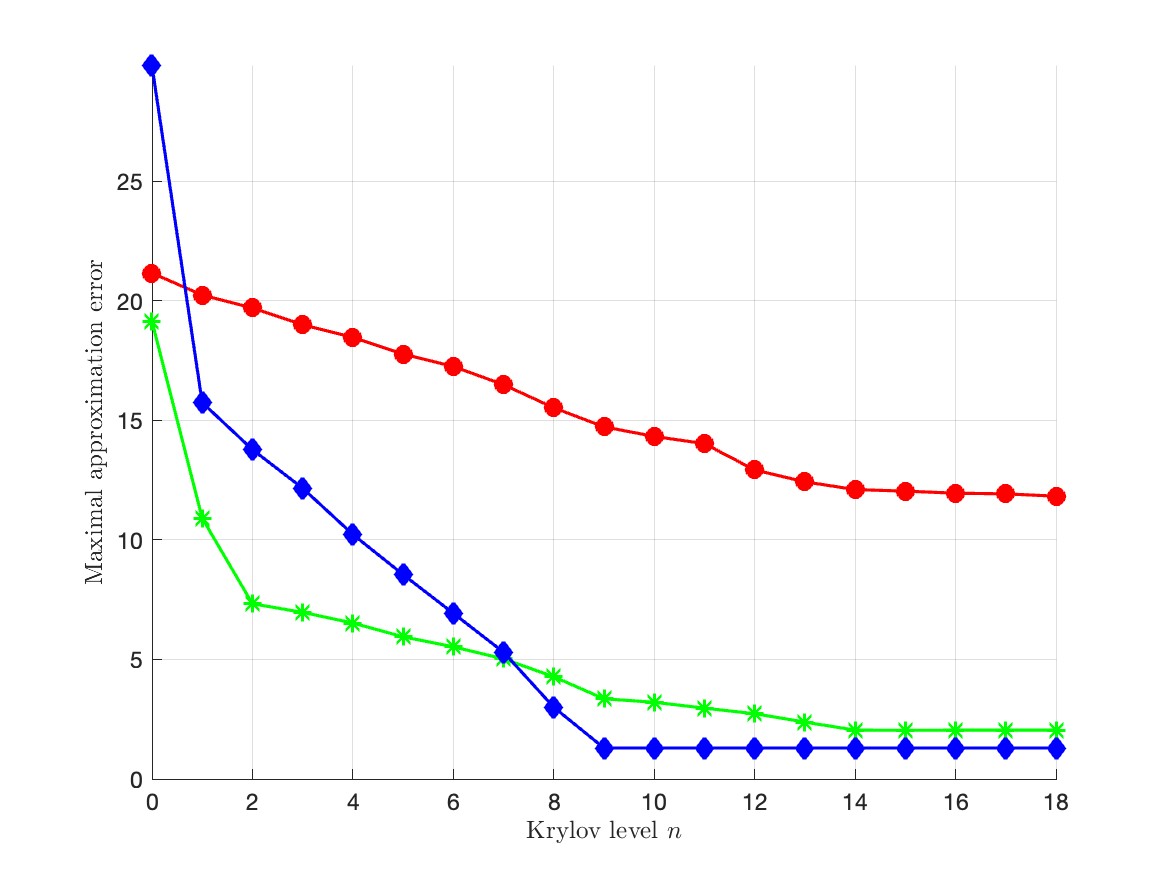}
\caption{Plotted on the top are the flight delay data ${\bf x}_d$  (left) and
 absolute error data $|{\bf x}_{K; d, n}-{\bf x}_d|$ (right) of the top 50 US airports on 29 August 2024, except
Honolulu International airport (HNL) and Luis Munoz Marin International airport (SJU), where $d=60$,
${\bf x}_{K; d, n}$ is reconstructed from Algorithm \ref{kylovsampling.alg}  in the noiseless environment
 with adaptive GSIS model and Krylov level $n=2$, and maximal approximation error is $\|{\bf x}_{K; 60, 2}-{\bf x}_{60}\|_\infty=7.9382$.
Shown on the bottom are the average of   maximal approximation error  $F_{K, n}$ by Krylov subspaces of adaptive GSIS  (in green) and of non-adaptive GSIS (in blue)
and the maximal approximation error    $F_{B, n}$ by bandlimited  spaces (in red) over 184 days, where the subsampling set contains the whole 50 airports  for the left plot
and the top 30 airports with most delay on average over 184 days for the right plot. }
\label{flightdelay.fig}
\end{figure}

  Flight delays
  are inevitable especially for frequent flyers.
In this subsection, we consider modeling 
arrival performance (measured as delays in minutes) of the top 50 airports  out of
$323$ airports in the USA with the highest traffic volume \cite{Romero2017, flightdata}, and select
the dataset
 on July, August, and September in 2014 and 2015 in our simulations for total 184 days, see the top left plot of
 Figure \ref{flightdelay.fig}.

We model the flight delay dataset as a family of graph signals  ${\bf x}_d=(x_d(n))_{n\in A}, 1\le d\le 184$, on the  
undirected graph ${\mathcal F}=(A, D)$, where
$d$ indicates the date in the dataset,
the vertex set $A$ has vertices indicating  the top 50 airports and
the edge set $D$ contains all airport pairs so that  the number of mutual flights between them exceeds $100$ in the three months 
(roughly one direct flight between them per day).
The underlying graph is designed to ensure that our analysis focuses on the busiest and most significant hubs, which are likely to have a greater influence on overall air traffic patterns and network dynamics. In the simulations, we use the corresponding symmetrically normalized graph Laplacian  ${\bf L}^{\rm sym}$ as the graph shift in the definition of   our proposed GSIS and bandlimiting model.

Three most common causes of flight delay are  extreme  weather condition, air traffic congestion and technical issues by  operating airlines. The first two sources may lead to delay of most flights leaving and landing at the airports in certain region, while the last reason may create a ripple effect for flights operated by the airline at its  hub airports.
In our simulations, we  consider nonadaptive  GSIS
$$H={\rm span}\{({\bf L}^{\rm sym})^n \delta_{i}\ | \  n\ge 0, 1\le i\le 3\}$$
and  adaptive GSISs
$$H_d={\rm span}\{({\bf L}^{\rm sym})^n \delta_{i, d}\ | \  n\ge 0, 1\le i\le 3\}, \ 1\le d\le 184, $$
to model graph signals  ${\bf x}_d, 1\le d\le 184$, where
$\delta_{i}$ and $\delta_{i, d}, 1\le i\le 3$, are delta signals at vertices representing airports with top three delay on average in the dataset
and at day $d, 1\le d\le 184$, respectively.
Let $H_n$ and $H_{d, n}, 0\le n\le 49$, be the corresponding Krylov spaces spanned by
$({\bf L}^{\rm sym})^m\delta_{i}$ and
$({\bf L}^{\rm sym})^m\delta_{i, d}$ with $0\le m\le n, 1\le i\le 3$ respectively.  Shown at
the bottom left plot of
 Figure \ref{flightdelay.fig} is  the average of (non)adataptive maximal Krylov approximation error
 and bandlimiting approximation error,
 \begin{equation}\label{modelsuitability.def} F_{K, n}={\|{\bf x}_d-{\bf x}_{K; d, n}\|_\infty}
  \ {\rm and} \ F_{B, n}={\|{\bf x}_d-{\bf x}_{B; d, n}\|_\infty}\end{equation}
over 184 days, where  ${\bf x}_{K; d, n}$ is the signal in $H_{d, n}$ (respectively $H_n$)
 reconstructed from Algorithm \ref{kylovsampling.alg} in the noiseless environment with the identity matrix as the sampling matrix,
 and ${\bf x}_{B; d, n}$ is the projection of the graph signal ${\bf x}_d$ on the bandlimiting space $B_{d, n}$ of lowest frequencies with
its dimension being the same
 of the one of  Kylov space, i.e.,  $\dim B_{d, n}=\dim H_{d, n}$.
We also test the performance of  Algorithm \ref{kylovsampling.alg} 
when  the flight delay  data on the 30 airports with top  delay on average are available only, see the bottom right plot of Figure \ref{flightdelay.fig}.

 The measurements  in \eqref{modelsuitability.def}
  can be used to measure the  rationality to use GSISs and bandlimited spaces to model
 the flight delay dataset. From the plots on the top right and at the bottom of Figure \ref{flightdelay.fig}, we see that
 the GSIS is more suitable to model flight delay dataset than the bandlimited space does, and that
 the GSIS with the generators chosen adaptively could further improve the performance.
 The reason could be that the flight delay data is more well-localized in spatial domain
than concentrated on low frequency in the Fourier domain.
In the United States, the words on time at the airport refer to any flight departure or flight arrival within less than 15 minutes of their scheduled time.
We observe that the maximal Krylov approximation error with level $n=2$ for the adaptive GSIS model is less than 9 minutes, which confirms the speculation that
the flight delays at the US  airport network are mostly caused by few airports and their subsequent ripple
effect. 
For the scenario that only flight delay data on the top 30 airports with most delays is available, the corresponding data fitting via the nonadaptive/adaptive GSIS models have comparable performance,  as we notice that the flight delay data  at airports representing the supporting vertices of the generators in the nonadaptive GSIS
are always available, while  no airports with the most delays on some day, used for the adaptive GSIS,
 are included in those 30 airports with the most delays on average.


\section{Proofs}
\label{proof.section}
In this section, we collect the proofs of Theorems \ref{sis.thm}, \ref{finitegenerated.thm},
\ref{SisRkhs.thm} and \ref{rkhs.thm}.

\vskip-.4in

\subsection{Proof of Theorem \ref{sis.thm}}\label{sis.thm.pfsection}
The implications (iv)$\Longrightarrow$(iii)$\Longrightarrow$ (ii) follow from the definition of a finite-generated shift-invariant space, and
the implication (i)$\Longrightarrow$ (iv) holds by Theorem \ref{finitegenerated.thm} with its proof given in Appendix  \ref{finitegenerated.thm.pfsection}.
Then it remains to prove the implication
(ii)$\Longrightarrow$(i).  

 For $1\le n\le N$, let ${\bf e}_n$ be the unit vector with zero entries except taking value one at $n$-th entries, and ${\bf E}_n$ be the diagonal matrix with ${\bf e}_n$
as the diagonal vector.   By Assumption \ref{graphshiftassumption}, there exist polynomials $p_n, 1\le n\le N$, of degree at most $N-1$ such that
\begin{equation} \label{sis.thm.pfeq1+}
p_n({\bf S}_1, \ldots, {\bf S}_L)= {\bf U} {\bf E}_n {\bf U}^T,\  1\le n\le N,
\end{equation}
where ${\bf U}$ is the orthogonal matrix in \eqref{shiftdiagonalization.def}.
In particular, the polynomials $p_n, 1\le n\le N$, are chosen to satisfying the  interpolation property
\begin{equation}  \label{sis.thm.pfeq2-} p_n({\pmb \Lambda}_{m})=\delta_{nm}, \ 1\le m,n\le N,
\end{equation}
where $\delta_{nm}$ is the standard Kronecker symbol \cite{approximationbook}.

For $1\le n\le N$, let ${\bf u}_n$ be the $n$-th column of the orthogonal matrix ${\bf U}$
in \eqref{shiftdiagonalization.def}, and set
\begin{equation}  \label{sis.thm.pfeq2}
G_n= p_n({\bf S}_1, \ldots, {\bf S}_L)H, \ 1\le n\le N.\end{equation}
Then one may verify that
\begin{equation} \label{sis.thm.pfeq2+}
G_n=\{ \widehat {\bf x}(n) {\bf u}_n\ \!| \ {\bf x}\in H\},\  1\le n\le N, \end{equation}
where $\widehat{\bf x}(n)={\bf e}_n^T \widehat{\bf x}$ is the $n$-th component of the Fourier transform $\widehat {\bf x}$ of a  signal ${\bf x}\in H$.
Therefore for any $1\le n\le N$, either $G_n$  is  trivial or it has dimension one, i.e.,
\begin{equation} \label{sis.thm.pfeq3}
 {\rm either}\ \
G_n=\{0\}\ \ {\rm  or} \ \ G_n={\rm span}\  {\bf u}_n.
\end{equation}

Let
$\Omega=\{n\ \!| \ G_n\ne \{0\}\}$, and set
\begin{equation} \label{sis.thm.pfeq4} B_\Omega=\{{\bf x} \ | \ {\rm supp}\ \widehat {\bf x}\subset \Omega\}=\bigoplus_{n\in \Omega} G_n.\end{equation}
By the shift-invariance assumption on $H$, we obtain from \eqref{sis.thm.pfeq2} that
$G_n\subset H, 1\le n\le N$, and hence
\begin{equation}  \label{sis.thm.pfeq6} B_\Omega\subset H\end{equation}
by \eqref{sis.thm.pfeq3} and \eqref{sis.thm.pfeq4}. On the other hand, for any ${\bf x}\in H$, we obtained from  \eqref{sis.thm.pfeq2}
and the definition of polynomials $p_n, 1\le n\le N$, that
$${\bf x}=\sum_{n=1}^N p_n({\bf S}_1, \ldots, {\bf S}_L) {\bf x}=\sum_{n\in \Omega} p_n({\bf S}_1, \ldots, {\bf S}_L) {\bf x}\in B_\Omega.$$
This together with \eqref{sis.thm.pfeq6} proves that
$H=B_\Omega$, 
and hence  completes the proof of the implication (ii)$\Longrightarrow$(i).

  \subsection{Proof of Theorem \ref{finitegenerated.thm}}\label{finitegenerated.thm.pfsection}

Let ${\pmb \lambda}(n), 1\le n\le N$, be as in \eqref{jointspectrum.def}.
 By Assumption \ref{graphshiftassumption}, there exists a nonzero vector
  ${\bf d}=[d_1, \ldots, d_L]^T\in {\mathbb R}^L$ such that
  ${\bf d}^T {\pmb \lambda}(n), 1\le n\le N$, are distinct real numbers.
Take ${\bf T}=\sum_{l=1}^L d_l {\bf S}_l$ and let
$\phi_0$ be as in \eqref{phi0.def}.  Then it remains to verify
that
\begin{equation}\label{BOmega.Hphi}
B_\Omega={\rm span} \{{\bf T}^m \phi_0, 0\le m\le \#\Omega-1\}.
\end{equation}

By \eqref{shift.Fourier} and \eqref{phi0.def}, we see that the Fourier transforms of
 ${\bf T}^m \phi_0,\ 0\le m\le \#\Omega-1$, are
  supported on $\Omega$ and hence
 \begin{equation} \label{BOmega.Hphi.eq1}
 {\rm span} \{{\bf T}^m \phi_0, 0\le m\le \#\Omega-1\}\subset B_\Omega.
 \end{equation}

Let
 $q_n, n\in \Omega$, be univariate polynomials of degree at most $\#\Omega-1$ that satisfy the  interpolation property
\begin{equation}  \label{finitegenerated.thm.pfeq2} q_n({\bf d}^T {\pmb \lambda}(n'))=\delta_{nn'},\ \ n, n'\in \Omega.
\end{equation}
The existence of such a polynomial follows from the distinct assumption on ${\bf d}^T {\pmb \lambda}(n'), n'\in \Omega$.
By the construction of the graph shift ${\bf T}$, we see that
${\bf T}={\bf U} {\pmb \Lambda}_T {\bf U}^T$ for some diagonal matrix
${\pmb \Lambda}_T$  with diagonal entries ${\bf d}^T {\pmb \lambda}(n'), 1\le n'\le N$.
Therefore
\begin{equation} \label{finitegenerated.thm.pfeq3}
q_n({\bf T})={\bf U} q_n({\pmb \Lambda}_T) {\bf U}^T\end{equation}
with the diagonal matrix $q_n({\pmb \Lambda}_T)$ having diagonal entries $q_n({\bf d}^T {\pmb \lambda}(n')), 1\le n'\le N$.

Take arbitrary bandlimited signal ${\bf x}\in B_\Omega$.
By \eqref{finitegenerated.thm.pfeq2} and \eqref{finitegenerated.thm.pfeq3}, we have
$$\widehat {\bf x}=\sum_{n\in \Omega} \frac{\widehat {\bf x}(n)}{\widehat \phi_0(n)}  q_n({\pmb \Lambda}_T) \widehat \phi_0.$$
Therefore taking the inverse Fourier transform ${\mathcal F}^{-1}$  at both sides yields
$${\bf x}=\sum_{n\in \Omega} \frac{\hat {\bf x}(n)}{\widehat \phi_0(n)} q_n({\bf T}) \phi_0.$$
This together with the degree property for the polynomials $q_n, n\in \Omega$ proves that
 \begin{equation} \label{BOmega.Hphi.eq2}
 B_\Omega \subset  {\rm span} \{{\bf T}^m \phi_0, 0\le m\le \#\Omega-1\}.
 \end{equation}

Combining \eqref{BOmega.Hphi.eq1} and \eqref{BOmega.Hphi.eq2} proves \eqref{BOmega.Hphi}
 and hence completes the proof of  Theorem \ref{finitegenerated.thm}.

\subsection{Proof of Theorem \ref{SisRkhs.thm}}
\label{SisRkhs.thm.proofsec}
 By Theorem \ref{sis.thm}, there exists $\Omega\subset \{1, \ldots, N\}$ such that
\eqref{bandlimit.def} holds.  Define
the kernel ${\bf K}$ by
\begin{equation}\label{SisRkhs.thm.pfeq1}
{\bf K}= {\bf U} {\pmb \Lambda}_\Omega {\bf U}^T,
\end{equation}
where ${\bf U}$ is the orthogonal matrix in \eqref{shiftdiagonalization.def} and
${\pmb \Lambda}_\Omega$ is the diagonal matrix in \eqref{standardinnerproduct.def}.
Then we obtain from \eqref{shiftdiagonalization.def} and \eqref{SisRkhs.thm.pfeq1} that
 the kernel ${\bf K}$ in \eqref{SisRkhs.thm.pfeq1} is shift-invariant.

 Let ${\pmb \delta}_j, j\in V$, be  delta signals taking value zero at all vertices except value one at the vertex $j$.
 By  \eqref{bandlimit.def} and \eqref{SisRkhs.thm.pfeq1},
  the linear space  $H$ is spanned by ${\bf K}{\pmb \delta}_j, j\in V$, and
  satisfies the following reproducing kernel property,
$${\bf x}={\bf K} {\bf x}, \ {\bf x}\in H,$$
where the standard Euclidean inner product on ${\mathbb R}^N$ is used for its linear subspace $H\subset {\mathbb R}^N$.
This completes the proof that $H$ is a reproducing kernel space with the shift-invariant space ${\bf K}$.

\subsection{Proof of Theorem \ref{rkhs.thm}}\label{rkhs.thm.pfsection}

 Let ${\bf K}$ be the shift-invariant kernel of the RKHS $H$, and
 let ${\bf x}, {\bf y}$ be  two arbitrary elements in $H$.
 By \eqref{SisRkhs.def2} and the definition of the inner product on the RKHS $H$,
\begin{equation}\label{rkhs.thm.pfeq2}
\langle {\bf x}, {\bf y}\rangle_H = {\bf c}^T {\bf K} {\bf d}.
\end{equation}
for some ${\bf c}$ and ${\bf d}\in {\mathbb R}^N$ with ${\bf x}={\bf K}{\bf c}$ and ${\bf y}={\bf K}{\bf d}$.

By  Theorem A.3 in \cite{ncjs22} and Assumption \ref{graphshiftassumption}, there exists a diagonal matrix $\pmb \Lambda$  for the
shift-invariant kernel  ${\bf K}$ such that
${\bf K}={\bf U} {\pmb \Lambda} {\bf U}^T$,
where ${\bf U}$ is the orthogonal matrix in \eqref{shiftdiagonalization.def}.
Denote the  pseudo-inverse of the diagonal matrix ${\pmb \Lambda}$ by ${\pmb \Lambda}^\dag$.
Then it follows from \eqref{rkhs.thm.pfeq2} that $\widehat {\bf x}= {\pmb \Lambda} \widehat {\bf c}, \widehat {\bf y}= {\pmb \Lambda} \widehat {\bf d}$
and
$$\langle {\bf x}, {\bf y}\rangle_H = {\widehat {\bf c}}^T {\pmb \Lambda} \widehat{\bf d}=
{\widehat {\bf c}}^T {\pmb \Lambda}   {\pmb \Lambda}^\dag  {\pmb \Lambda} \widehat{\bf d}
= {\widehat {\bf x}}^T {\pmb \Lambda}^\dag \widehat {\bf y}.$$
As ${\bf K}$ is a reproducing kernel,  it is represented by  a positive semidefinite matrix, which implies that
${\pmb \Lambda}^\dag$ has nonnegative entries.  Therefore \eqref{rkhs.thm.eq1} holds and the proof is completed. 

\end{document}